\documentclass[showkeys,preprintnumbers,amsmath,amssymb,nofootinbib]{revtex4}
\usepackage{graphicx}
\usepackage{dcolumn}
\usepackage{bm}
\usepackage{amsmath,amssymb,epsfig} 
\usepackage{wrapfig}

\begin{document}


\title{Experimental Tests of General Relativity:\\
Recent Progress and Future Directions}

\author{Slava G. Turyshev
}   
\affiliation{ 
Sternberg Astronomical Institute, 13 Universitetskij Prospect, 119992 Moscow, Russia
} 
\affiliation{ 
Jet Propulsion Laboratory, California Institute of Technology,\\
4800 Oak Grove Drive, Pasadena, CA 91109-0899, USA
}%

\date{\today}

\begin{abstract}  
Einstein's general theory of relativity is the standard theory of gravity, especially where the needs of astronomy, astrophysics, cosmology and fundamental physics are concerned. As such, this theory is used for many practically important purposes involving spacecraft navigation, geodesy, time transfer, etc. Here I review the foundations of general relativity, discuss recent progress in the tests of relativistic gravity, and present motivations for the new generation of high-accuracy tests of new physics beyond general relativity. Space-based experiments in fundamental physics are capable today to uniquely address important questions related to the fundamental laws of nature. I discuss the advances in our understanding of fundamental physics that are anticipated in the near future and evaluate the discovery potential of a number of the recently proposed space-based gravitational experiments.
\end{abstract}

\keywords  
{Tests of general theory of relativity; Standard Model extensions; cosmology; modified gravity; string theory; scalar-tensor theories; Equivalence Principle; gravitational experiments in space}

\maketitle


\section{\label{sec:intro}Introduction}

November 25, 2015 will mark the Centennial of general theory of relativity, which was developed by Albert Einstein between 1905 and 1915 \cite{Frederiks:1999,Vizgin:2001}. Ever since its original publication \cite{Einstein-1907,Einstein-1915,Einstein-1916}, the theory continues to be an active area of both theoretical and experimental research \cite{Turyshev-etal-2007,Turyshev:2008dr}.

The theory first demonstrated its empirical success in 1915 by explaining the anomalous perihelion precession of Mercury's orbit.  This anomaly was known long before Einstein; it amounts to 43 arcsec per century (''/cy) and cannot be explained within Newton's gravity, thereby presenting a challenge for physicists and astronomers.  In 1855, Urbain LeVerrier, who in 1846 predicted the existence of Neptune, a planet on an extreme orbit, wrote that the anomalous residue of the Mercurial precession would be accounted for if yet another planet, the as-yet undiscovered planet Vulcan, revolves inside the Mercurial orbit. Because of the proximity of the sun Vulcan would not be easily observed, but LeVerrier thought he, nevertheless, had detected it.  However, no confirmation came in the decades that followed. It took another 60 years to solve this puzzle. In 1915, before publishing the historical paper containing the field equations of general relativity (e.g. \cite{Einstein-1915}), Einstein computed the expected perihelion precession of Mercury's orbit.  When he obtained the famous 43 ''/cy needed to account for the anomaly, he realized that a new era in gravitational physics had just begun!  

Shortly thereafter, Sir Arthur Eddington's 1919 observations of star lines-of-sight during a solar eclipse \cite{Dyson-Eddington-Davidson-1920} confirmed the doubling of the deflection angles predicted by general relativity as compared to Newtonian and equivalence principle (EP) arguments.\footnote{Eddington had been aware of several alternative predictions for his experiment.  In 1801 Johann Georg von Soldner \cite{Lenard-1921} had pointed out that Newtonian gravity predicts that the trajectory of starlight bends in the vicinity of a massive object, but the predicted effect is only half the value predicted by general relativity as calculated by Einstein \cite{Einstein-1911}. Other investigators claimed that gravity will did affect light propagation.  Eddington's experiment settled these claims by pronouncing general relativity a winner.}  Observations were made simultaneously in the city of Sobral in Brazil and on the island of Principe off the west coast of Africa; these observations focused on determining the change in position of stars as they passed near the Sun on the celestial sphere.  The results were presented on November 6, 1919 at a special joint meeting of the Royal Astronomical Society and the Royal Society of London \cite{Coles-2001}.  The data from Sobral, with measurements of seven stars in good visibility, yielded deflections of $1.98\pm0.16$~arcsec.  The data from Principe were less convincing.  Only five stars were reliably measured, and the conditions there led to a much larger error.  Nevertheless, the obtained value was $1.61\pm0.4$~arcsec.  Both were within $2\sigma$ of Einstein's value of 1.74 and were more than two standard deviations away from both zero and the Newtonian value of 0.87. These observations became the first dedicated experiment to test general theory of relativity.\footnote{The early accuracy, however, was poor.  Dyson et al. \cite{Dyson-Eddington-Davidson-1920} quoted an optimistically low uncertainty in their measurement, which was argued by some to have been plagued with systematic error and possibly confirmation bias, although modern re-analysis of the dataset suggests that Eddington's analysis was accurate \cite{Kennefick-2007}. However, considerable uncertainty remained in these measurements for almost 50 years, until first interplanetary spacecraft and microwave tracking techniques became available. It was not until the late 1960s that it was definitively shown that the angle of deflection was the full value predicted by general relativity.} 
In Europe, which was still recovering from the World War I, this result was considered spectacular news and occupied the front pages of most major newspapers making general relativity an instant success. 

\begin{wrapfigure}{R}{0.50\textwidth}
  \vspace{-10pt}
  \begin{center}
    \includegraphics[width=0.50\textwidth]{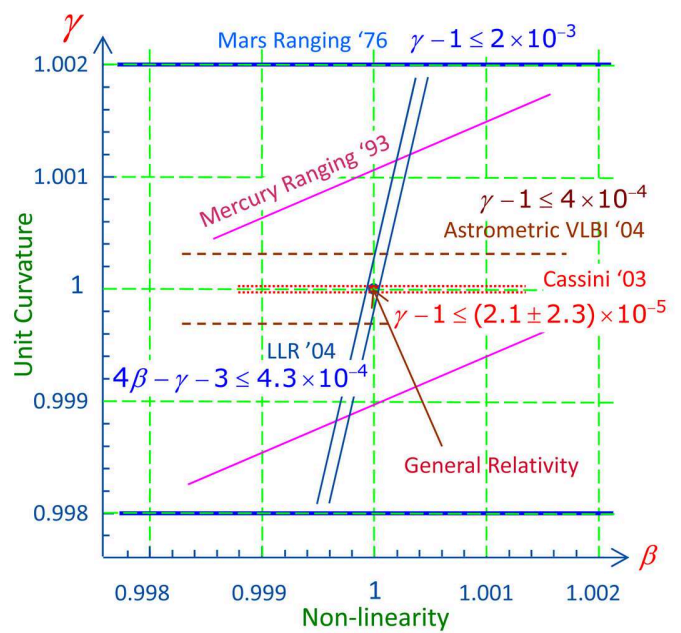}
  \end{center}
  \vspace{-10pt}
  \caption{The progress in improving the knowledge of the Eddington's parameters $\gamma$ and $\beta$ for the last 39 years (i.e., since 1969 \cite{Williams-etal-2005}). So far, general theory of relativity survived every test \cite{Turyshev:2008dr}, yielding $\gamma-1=(2.1\pm2.3)\times10^{-5}$ \cite{Bertotti-Iess-Tortora-2003} and $\beta-1=(1.2\pm1.1)\times10^{-4}$ \cite{Williams-etal-2004}.}
\label{fig:gamma-beta}
  \vspace{-10pt}
\end{wrapfigure}

From these beginnings, the general theory of relativity has been verified at ever higher accuracy; presently, it successfully accounts for all data gathered to date. The true renaissance in the tests of general relativity began in 1970s with major advances in several disciplines notably in microwave spacecraft tracking, high precision astrometric observations, and  lunar laser ranging (LLR) (see Fig.\ref{fig:gamma-beta}). 

Thus, analysis of 14 months's worth of data obtained from radio ranging to the Viking spacecraft verified, to an estimated accuracy of 0.1\%, the prediction of the general theory of relativity that the round-trip times of light signals traveling between the earth and Mars are increased by the direct effect of solar gravity \cite{viking_shapiro1,viking_shapiro2,viking_reasen}. The corresponding value for the Eddington's metric parameter\footnote{To describe the accuracy achieved in the solar system experiments, it is useful to refer to the parameterized post-Newtonian (PPN) formalism (see discussion in Sec.~\ref{sec:ppn} and \cite{Will_book93}). Two parameters are of interest here, the PPN parameters $\gamma$ and $\beta$, whose values in general relativity are $\gamma = \beta=1$. The introduction of $\gamma$ and $\beta$ is useful with regard to measurement accuracies \cite{Turyshev:1996}. In the PPN formalism (originally developed by Eddington), the angle of light deflection is proportional to ($\gamma$ + 1)/2, so that astrometric measurements might be used for a precise determination of the parameter $\gamma$. The parameter $\beta$ contributes to the relativistic perihelion precession of a body's orbit (see \cite{Will_book93}).}. In 1978 the value of the PPN parameter $\gamma$ was obtained at the level of $1.000 \pm 0.002$. Spacecraft and planetary radar observations have reached an accuracy of $\sim$0.15\%  \cite{Williams_etal_2001,Anderson_etal_02,Pitjeva-2005}.  

Meanwhile, very long baseline interferometry (VLBI) has achieved accuracies of better than 0.1 mas (milliarcseconds of arc), and regular geodetic VLBI measurements have frequently been used to determine the space curvature parameter $\gamma$. Detailed analyses of VLBI data have yielded a consistent stream of improvements $\gamma= 1.000\pm0.003$ \cite{Robertson-Carter-1984,RoberstonCarterDillinger91}, $\gamma = 0.9996 \pm 0.0017$ \cite{Lebach95}, $\gamma = 0.99994 \pm 0.00031$ \cite{Eubanks97} and $\gamma=0.99983\pm0.00045$ \cite{Shapiro_SS_etal_2004}, resulting in the accuracy of better than $\sim$0.045\% in the tests of gravity via astrometric VLBI observations.

LLR, a continuing legacy of the Apollo program, provided improved constraint on the combination of parameters $4\beta-\gamma-3$ \cite{JimSkipJean96,Williams_etal_2001,Williams-Turyshev-Murphy-2004,Williams-etal-2004,Williams-etal-2005}
In 2004, the analysis of LLR data \cite{Williams-etal-2004} constrained this combination as $4\beta-\gamma-3=(4.0\pm4.3)\times10^{-4}$, leading to an accuracy of $\sim$0.011\% in verification of general relativity via precision measurements of the lunar orbit.  

Finally, microwave tracking of the Cassini spacecraft on its approach to Saturn improved the measurement accuracy of the parameter $\gamma$ to $\gamma-1=(2.1\pm2.3)\times10^{-5}$, thereby reaching the current best accuracy of $\sim$0.002\% provided by tests of gravity in the solar system \cite{Iess_etal_1999,Bertotti-Iess-Tortora-2003}. 

To date, general relativity is also in agreement with data from the binary and double pulsars. Recently, investigators have shown a considerable interest in the physical processes occurring in the strong gravitational field regime with relativistic pulsars providing a promising possibility to test gravity in this qualitatively different dynamical environment. Although, strictly speaking, the binary pulsars move in the weak gravitational field of a companion, they do provide precision tests of the strong-field regime \cite{Lange_etal_2001}. This becomes clear when considering strong self-field effects which are predicted by the majority of alternative theories. Such effects would, clearly affect the pulsars' orbital motion, allowing us to search for these effects and hence providing us with a unique precision strong-field test of gravity. The general theoretical framework for pulsar tests of strong-field gravity was introduced in \cite{DamourTaylor92}; the observational data for the initial tests were obtained with PSR1534 \cite{Taylor_etal92}. An analysis of strong-field gravitational tests and their theoretical justification is presented in \cite{Damour-EspFarese-1996-1,Damour-EspFarese-1996-2,Damour_EspFarese98}. By measuring relativistic corrections to the Keplerian description of the orbital motion, the recent analysis of the data collected from the double pulsar system, PSR J0737-3039A/B, found agreement with the general relativity within an uncertainty of $\sim 0.05\%$ at a $3 \sigma$ confidence level \cite{Kramer-etal-2006}, to date the most precise pulsar test of gravity yet obtained. 

As a result, both in the weak field limit (as in our solar system) and with the stronger fields present in systems of binary pulsars the predictions of general relativity have been extremely well tested. It is remarkable that more than 90 years after general relativity was conceived, Einstein's theory has survived every test \cite{Will-lrr-2006-3}.  Such longevity and success make general relativity the de-facto ``standard'' theory of gravitation for all practical purposes involving spacecraft navigation and astrometry, astronomy, astrophysics, cosmology and fundamental physics \cite{Turyshev-etal-2007}. 

However, despite its remarkable success, there are many important reasons to question the validity of general relativity and to determine the level of accuracy at which it is violated.  
On the theoretical front, problems arise from several directions, most concerning the strong-gravitational field regime. This challenges include the appearance of spacetime singularities and the inability of classical description to describe the physics of very strong gravitational fields. A way out of this difficulty may be through gravity quantization. However, despite the success of modern gauge-field theories in describing the electromagnetic, weak, and strong interactions, we do not yet understand how gravity should be described at the quantum level. 

The continued inability to merge gravity with quantum mechanics along with recent cosmological observations, indicates that the pure tensor gravity of general relativity needs modification.  In theories that attempt to include gravity, new long-range forces arise as an addition to the Newtonian inverse-square law. Regardless of whether the cosmological constant should be included, there are also important reasons to consider additional fields, especially scalar fields. Although the latter naturally appear in these modern theories, their inclusion predicts a non-Einsteinian behavior of gravitating systems. These deviations from general relativity lead to violation of the EP, and to modification of large-scale gravitational phenomena, and they cast doubt upon the constancy of the fundamental constants. These predictions motivate new searches for very small deviations of relativistic gravity from the behavior prescribed by  general relativity; they also provide a new theoretical paradigm and guidance for future space-based gravity experiments \cite{Turyshev-etal-2007,Turyshev:2008dr}.

\begin{figure}[h!]
  \vspace{-0pt}
  \begin{center}
    \includegraphics[width=0.95\textwidth]{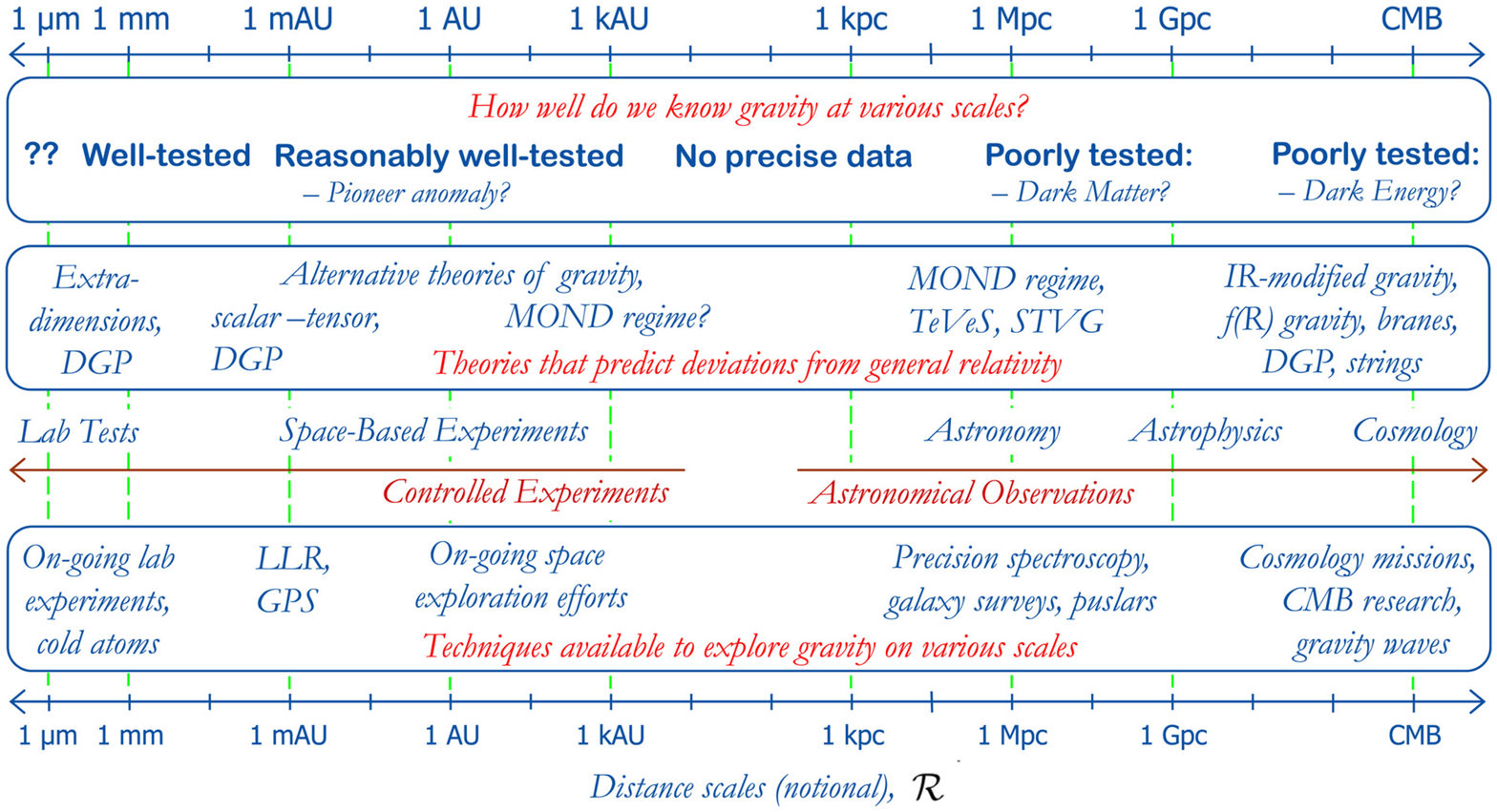}
  \end{center}
  \vspace{-10pt}
  \caption{Present knowledge of gravity at various distance scales. Given the recent impressive progress in many measurements technologies \cite{Turyshev:2008dr}, solar system experiments offer great potential to improve our knowledge of relativistic gravity. Abbreviations: CMB, cosmic microwave background; DGP, Dvali-Gabadadze-Porrati model \cite{Dvali-Gabadadze-Porrati-2000}; GPS, Global Positioning System; LLR, lunar laser ranging; MOND, modified Newtonian dynamics \cite{Milgrom01}; TeVeS, tensor-vector-scalar gravity \cite{Bekenstein:2004ne}; STVG, scalar-tensor-vector-gravity \cite{Moffat:2005si}; IR-modified gravity (for review, see \cite{Rubakov:2008nh}); $f(R)$ gravity (for review, see \cite{Sotiriou:2008rp}).}
\label{fig:gravity-knowledge}
  \vspace{-10pt}
\end{figure}

Note that on the largest spatial scales, such as the galactic and cosmological scales, general relativity has not yet been subject to precision tests.  Some researchers have interpreted observations supporting the presence of dark matter and dark energy as a failure of general relativity at large distances, at small accelerations, or at small curvatures (see discussion in Refs.~\cite{Turyshev:2008dr,Milgrom01,Bekenstein:2004ne,Moffat:2005si,Dvali-Gabadadze-Porrati-2000,Rubakov:2008nh,Sotiriou:2008rp}). Fig.~\ref{fig:gravity-knowledge} shows our present knowledge of gravity at various distance scales; it also indicates the theories that have been proposed to explain various observed phenomena and the techniques that have been used to conduct experimental studies of gravity at various regimes. The very strong gravitational fields that must be present close to black holes, especially those supermassive black holes that are thought to power quasars and less active active galactic nuclei, belong to a field of intensely active research. Observations of these quasars and active galactic nuclei are difficult to obtain, and the interpretation of the observations is heavily dependent upon astrophysical models other than general relativity and competing fundamental theories of gravitation; however, such interpretations are qualitatively consistent with the black-hole concept as modeled in general relativity.

Today physics stands at the threshold of major discoveries \cite{Witten:2001,Turyshev:2008dr,Wilczek:2007gsa}. Growing observational evidence points to the need for new physics. Efforts to discover new fundamental symmetries, investigations of the limits of established symmetries, tests of the general theory of relativity, searches for gravitational waves, and attempts to understand the nature of dark matter were among the topics that had been the focus of the scientific research at the end of the last century.  These efforts have further intensified with the discovery of dark energy made in the late 1990s, which triggered many new activities aimed at answering important questions related to the most fundamental laws of Nature \cite{Wilczek:2006,Wilczek:2007gsa,Turyshev-etal-2007}.

Historically, the nature of matter on Earth and the laws governing it were discovered in laboratories on Earth.  Scientific progress in these experiments depends on clever experimental strategy and the use of advanced technologies needed to overcome the limits imposed by the environment typically present in Earth-based laboratories.  Since recently, in many cases the conditions in these laboratories cannot be improved to the required levels of purity, thus, a space deployment becomes a natural and well justified next step.

Space deployment offers access to the conditions with a particular dynamical ``purity'' that are not available on Earth, but are pivotal for many pioneering investigations exploring the limits of modern physics. In particular, for many fundamental physics experiments, especially those aiming at exploring gravitation, cosmology, atomic physics, while achieving uttermost measurement precision, and increasingly so for high energy and particle astrophysics, a space-based location is the ultimate destination.

There are two approaches to physics research in space: one can detect and study signals from remote astrophysical objects (the ``observatory'' mode) or one can perform carefully designed experiments in space (the ``laboratory'' mode). Fig.~\ref{fig:gravity-knowledge} emphasizes the ``areas of responsibility'' of these two disciplines - observational and space-based laboratory  research in fundamental physics.  The two methods are complementary and the latter, which is the focus of this paper, has the advantage of utilizing the well-understood and controlled environments of a space-based laboratory. 

Considering gravitation and fundamental physics, our solar system is the laboratory that offers many opportunities to improve the tests of relativistic gravity. A carefully designed gravitational experiment has the advantage to conduct tests in a controlled and  well-understood environment and can achieve accuracies superior to its ground-based counterpart. Existing technologies allow one to take advantage of the unique environments found only in space, including variable gravity potentials, large distances, high velocity and low acceleration regimes, availability of pure geodetic trajectories, microgravity and thermally-stable environments (see discussion in \cite{Turyshev-etal-2007}).  Among the favorable factors are also the greatly reduced contribution of non-gravitational sources of noise (presently down to $10^{-14} ~{\rm m s}^{-2}{\rm Hz}^{-1/2}$), access to significant variations of gravitational potential (thus, compared with terrestrial conditions, the Sun offers a factor of $\sim$3,000 increase in the strength of gravitational effects) and corresponding gravitational acceleration (for instance, a highly eccentric solar orbit with apoapsis of 5 AU and periapsis of 10 solar radii offers more than 2 orders of magnitude in variation of the solar gravity potential and 4 orders of magnitude in variation in the corresponding gravitational acceleration, clearly not available otherwise), and also access to large distances, velocities, separations, availability of remote benchmarks and inertial references --- the conditions that simply can not be achieved on the ground. 

With recent advances in several applied physics disciplines new instruments and technologies have become available. These include highly accurate atomic clocks, optical frequency combs, atom interferometers, drag-free technologies, low-thrust micro-propulsion techniques, optical transponders, long-baseline optical interferometers, etc. \cite{Turyshev-q2c:2006,Phillips-2007}. Today, a new generation of high performance quantum sensors (ultra-stable atomic clocks, accelerometers, gyroscopes, gravimeters, gravity gradiometers, etc.) is surpassing previous state-of-the-art instruments, demonstrating the high potential of these techniques based on the engineering and manipulation of atomic systems.  Atomic clocks and inertial quantum sensors represent a key technology for accurate frequency measurements and ultra-precise monitoring of accelerations and rotations (see discussion in \cite{Turyshev-etal-2007}).  

New quantum devices based on ultra-cold atoms will enable fundamental physics experiments testing quantum physics, physics beyond the Standard Model of fundamental particles and interactions, special relativity, gravitation and general relativity.
Some of these instruments are already space-qualified, thereby enabling a number of high-precision investigations in laboratory fundamental physics in space. Therefore, in conjunction with the new high-precision measurement technologies, the unique conditions of space deployment are critical for the progress in the gravitational research.  As a result, modern-day space-based experiments are capable of reaching very high accuracies in testing the foundations of modern physics and are well positioned to provide  major advances in this area \cite{Turyshev:2008dr}.  

In this paper I discuss recent solar-system gravitational experiments that have contributed to the progress in relativistic gravity research by providing important guidance in the search for the next theory of gravity.  I also present theoretical motivation for new generation of high-precision gravitational experiments and discuss a number of recently proposed space-based tests of relativistic gravity.

The paper is organized as follows. Section~\ref{sec:gr-survey} discusses the foundations of the general theory of relativity and reviews the results of recent experiments designed to test the  foundations of this theory. We present the parameterized post-Newtonian formalism -- a phenomenological framework that is used to facilitate experimental tests of relativistic gravity.  Section~\ref{sec:beyond} presents  motivations for extending the theoretical model of gravity provided  by general relativity; it presents models arising from string theory,  discusses the scalar-tensor theories of gravity, and highlights the phenomenological implications of these proposals. I briefly review recent proposals to modify gravity on large scales and review their experimental implications. Section \ref{sec:future-tests} discusses future space-based experiments that aim to expand our knowledge of gravity. I focus on the space-based tests of general theory of relativity and discuss the experiments aiming to test the EP, local Lorentz and position invariances, the search for variability of the fundamental constants, tests of the gravitational inverse-square law and tests of alternative and modified-gravity theories. I present a list of the proposed space missions, focusing only on the most representative and viable concepts. I conclude in Section \ref{sec:conclusions}.

\section{Testing The Foundations of General Relativity}
\label{sec:gr-survey}

General relativity is a tensor field theory of gravitation with universal coupling to the particles and fields of the Standard Model. It describes gravity as a universal deformation of the flat spacetime Minkowski metric, $\gamma_{mn}$:
\begin{equation}
g_{mn}(x^k)=\gamma_{mn}+h_{mn}(x^k).
\end{equation}
Alternatively, it can also be defined as the unique, consistent, local theory of a massless spin-2 field $h_{mn}$, whose source is the total, conserved energy-momentum tensor (see \cite{Damour-2006-pdg} and references therein).  

Classically \cite{Einstein-1915,Einstein-1916}, the general theory of relativity is defined by two postulates.  One of the postulates states that the action describing the propagation and self-interaction of the gravitational field is given by
\begin{eqnarray}
\label{eq:hilb-ens-action}
{\cal S}_{\rm G}[g_{mn}]
&=&\frac{c^4}{16\pi G_{\rm N}}\int d^4x
\sqrt{-g} R,
\end{eqnarray}
where $G_{\rm N}$ is Newton's universal gravitational constant, $g^{mn}$ is the matrix inverse of $g_{mn}$, and $g=\det g_{mn}$, $R$ is the Ricci scalar given as $R=g^{mn}R_{mn}$ with the quantity $R_{mn}=\partial_k\Gamma^k_{mn}-\partial_m\Gamma^k_{nk}+\Gamma^k_{mn}\Gamma^l_{kl}-\Gamma^k_{ml}\Gamma^l_{nk}$ being the Ricci tensor and $\Gamma^k_{mn}=\frac{1}{2}g^{kp}(\partial_m g_{pn}+\partial_n g_{pm}-\partial_p g_{mn})$ are the Christoffel symbols.

The second postulate states that $g_{mn}$ couples universally, and minimally, to all the fields of the Standard Model by replacing the Minkowski metric everywhere. Schematically (suppressing matrix indices and labels for the various gauge fields and fermions and for the Higgs doublet), this postulate can be given by
{} 
\begin{eqnarray}
{\cal S}_{\rm SM}[\psi, A_m, H; g_{mn}]
&=&\int d^4x\Big[-\frac{1}{4}\sum \sqrt{-g} 
g^{mk} g^{nl}F^{\rm a}_{mn}F^{\rm a}_{kl} -
\sum \sqrt{-g}{\bar \psi}\gamma^m D_m\psi \nonumber\\
&&~\,~~~~~~~~- \frac{1}{2}\sqrt{-g}g^{mn}\overline{D_m H}D_n H-
\sqrt{-g} V(H)-\sum \lambda \sqrt{-g}{\bar \psi} H \psi -\sqrt{-g}\rho_{\rm vac}\Big],
\label{eq:action-SM}
\end{eqnarray}
where $\gamma^m \gamma^n + \gamma^n \gamma^m = 2 g^{mn}$, the covariant derivative $D_m$ contains, besides the usual gauge field terms, a (spin-dependent) gravitational contribution $\Gamma_m(x)$ \cite{Weinberg_book72} and $\rho_{\rm vac}$ is the vacuum energy density.
Applying the variational principle with respect to $g_{mn}$ to the total action 
\begin{equation}
{\cal S}_{\rm tot}[\psi,A_m,H;g_{mn}]={\cal S}_{\rm G}[g_{mn}]+{\cal S}_{\rm SM}[\psi, A_m, H; g_{mn}],
\end{equation}
one obtains the well-known Einstein's field equations of the general theory of relativity,
\begin{equation}
R_{mn}- \frac{1}{2} g_{mn} R +\Lambda g_{mn}= \frac{8\pi G_N}{c^4}T_{mn},
\label{eq:Einstein-eq}
\end{equation}
where $T_{mn}=g_{mk}g_{nl}T^{kl}$ with $T^{mn}=2/\sqrt{-g}~\delta {\cal L}_{\rm SM}/\delta g_{mn}$ being the (symmetric) energy-momentum tensor of the matter as described by the Standard Model with the Lagrangian density ${\cal L}_{\rm SM}$. With the value for the vacuum energy density $\rho_{\rm vac} \approx (2.3 \times 10^{-3} eV)^4$, as measured by recent cosmological observations \cite{Weinberg-1989:RevModPhys,Spergel-etal-2006}, the cosmological constant $\Lambda = 8\pi G_N \rho_{\rm vac}/c^4$ is too small to be observed by solar system experiments, but is clearly important for grater scales. 

The theory is invariant under arbitrary coordinate transformations: $x^{\prime m}=f^m(x^n)$. To solve the field equations Eq.~(\ref{eq:Einstein-eq}), one needs to fix this coordinate gauge freedom. E.g., the ``harmonic gauge'' (which is the analogue of the Lorentz gauge, $\partial_m A^m = 0$, in electromagnetism) corresponds to imposing the condition $\partial_n \sqrt{-g}g^{mn} = 0$.

Einstein's equations Eq.~({\ref{eq:Einstein-eq}) link the geometry of a four-dimensional, Riemannian manifold representing space-time with the energy-momentum contained in that space-time. Phenomena that, in classical mechanics, are ascribed to the action of the force of gravity (such as free-fall, orbital motion, and spacecraft trajectories), correspond to inertial motion within a curved geometry of spacetime in general relativity. 

\subsection{Scalar-Tensor Extensions to General Relativity}
\label{sec:scalar-tensor}

Metric theories have a special place among alternative theories of gravity. The reason for this is that, independently of the different principles at their foundations, the gravitational field in these theories affects the matter directly through the metric tensor $g_{mn}$, which is determined from the particular theory's field equations. As a result, in contrast to Newtonian gravity, this tensor expresses the properties of a particular gravitational theory and carries information about the gravitational field of the bodies.

In many alternative theories of gravity, the gravitational coupling strength exhibits a dependence on a field of some sort; in scalar-tensor theories, this is a scalar field $\varphi$. A general action for these theories can be written as
{}
\begin{eqnarray}  
S&=&{c^3\over 4\pi G}\int d^4x \sqrt{-g}
\left[\frac{1}{4}f(\varphi) R - \frac{1}{2}g(\varphi) \partial_{\mu} \varphi
\partial^{\mu} \varphi + V(\varphi) \right]
+\sum_{i}
q_{i}(\varphi)\mathcal{L}_{i}, \label{eq:sc-tensor} 
\end{eqnarray}
\noindent where $f(\varphi)$, $g(\varphi)$, and $V(\varphi)$ are generic functions, $q_i(\varphi)$ are coupling functions, and $\mathcal{L}_{i}$ is the Lagrangian density of the matter fields of the Standard Model Eq.~(\ref{eq:action-SM}). 

The Brans--Dicke theory \cite{Brans-Dicke-1961} is the best known alternative theory of gravity. It corresponds to the choice
{}
\begin{equation} 
\label{eq6:2} 
f(\varphi) = \varphi, \qquad g(\varphi) =
{\omega \over \varphi}, \qquad V(\varphi)=0.
\end{equation}
\noindent
Note that in the Brans--Dicke theory the kinetic energy term of the field $\varphi$ is non-canonical and that the latter has a dimension of energy squared. In this theory, the constant $\omega$ marks observational deviations from general relativity, which is recovered in the limit $\omega \to \infty$. In the context of the Brans-Dicke theory, one can operationally introduce Mach's Principle which states that the inertia of bodies is due to their interaction with the matter distribution in the Universe. Indeed, in this theory the gravitational coupling is proportional to $\varphi^{-1}$, which depends on the energy-momentum tensor of matter through the field equations. 
The stringent observational bound resulting from the 2003 experiment with the Cassini spacecraft require that $|\omega| \gtrsim 40\,000$ \cite{Bertotti-Iess-Tortora-2003,Will-lrr-2006-3}. There exist additional alternative theories that provide guidance for gravitational experiments (see \cite{Will-lrr-2006-3} for review).

\subsection{Metric Theories of Gravity and PPN Formalism}
\label{sec:ppn}

Generalizing on a phenomenological parameterization of the gravitational metric tensor field, which Eddington originally developed for a special case, a method called the parameterized post-Newtonian (PPN) formalism has been developed \cite{Nordtvedt_1968a,Nordtvedt_1968b,Nordtvedt_1969,Thorne-Will-1971-I,Thorne-Will-1971-II,Thorne-Will-1971-III,Will-1971-I,Will-1971-II,Will-Nordtvedt-72-I,Will-Nordtvedt-72-II,Will-1973,Will_book93}. This method represents the gravity tensor's potentials for slowly moving bodies and weak inter-body gravity, and is valid for a broad class of metric theories, including general relativity as a unique case. The several parameters in the PPN metric expansion vary from theory to theory, and they are individually associated with various symmetries and invariance properties of the underlying theory  (see \cite{Will_book93} for details).

If (for the sake of simplicity) one assumes that Lorentz invariance, local position invariance and total momentum conservation hold, the metric 
tensor for a system of $N$ point-like gravitational sources
in four dimensions may be written as 
{}
\begin{eqnarray}
\label{eq:metric}
g_{00}&=&1-\frac{2}{c^2}\sum_{j\not=i}\frac{\mu_j}{r_{ij}}+
\frac{2\beta}{c^4}\Big[\sum_{j\not=i}\frac{\mu_j}{r_{ij}}\Big]^2-
\frac{1+2\gamma}{c^4}\sum_{j\not=i}\frac{\mu_j{\dot r}^2_j}{r_{ij}}+
\frac{2(2\beta-1)}{c^4}\sum_{j\not=i}\frac{\mu_j}{r_{ij}}
\sum_{k\not=j}\frac{\mu_k}{r_{jk}}-
\frac{1}{c^4}\sum_{j\not=i}\mu_j\frac{\partial^2 r_{ij}}{\partial t^2}+
{\cal O}(c^{-5}),\nonumber\\
g_{0\alpha}&=& \frac{2(\gamma+1)}{c^3}\sum_{j\not=i}\frac{\mu_j{\dot {\bf r}}^\alpha_j}{r_{ij}}+
{\cal O}(c^{-5}),\\\nonumber
g_{\alpha\beta}&=&-\delta_{\alpha\beta}\Big(1+\frac{2\gamma}{c^2}\sum_{j\not=i}\frac{\mu_j}{r_{ij}}
+\frac{3\delta}{2c^4}\Big[\sum_{j\not=i}\frac{\mu_j}{r_{ij}}\Big]^2
\Big)+{\cal O}(c^{-5}),   
\end{eqnarray}

\noindent where the indices $j$ and $k$ refer to the $N$ bodies and where $k$ includes body $i$, whose motion is being investigated. $\mu_j$ is the gravitational constant for body $j$ given as $\mu_j=Gm_j$, where $G$ is the universal Newtonian gravitational constant and $m_j$ is the isolated rest mass of a body $j$. In addition, the vector ${\bf r}_i$ is the barycentric radius-vector of this body, the vector ${\bf r}_{ij}={\bf r}_j-{\bf r}_i$ is the
vector directed from body $i$ to body  $j$, $r_{ij}=|{\bf r}_j-{\bf r}_i|$, and the vector ${\bf n}_{ij}={\bf r}_{ij}/r_{ij}$ is the unit vector along this direction. 

Although general relativity replaces the scalar gravitational potential of classical physics by a symmetric rank-two tensor, the latter reduces to the former in certain limiting cases: For weak gravitational fields and slow speed (relative to the speed of light), the theory's predictions converge on those of Newton's law of gravity with some post-Newtonian corrections. The $1/c^2$ term in $g_{00}$ is the Newtonian limit; the $1/c^4$ terms multiplied by the parameters $\beta $ and $\gamma$,  are post-Newtonian terms. 
The term multiplied by the post-post-Newtonian parameter  $\delta$ also enters the calculation of the relativistic light propagation for some modern-day experiments \cite{Turyshev-etal-2007,Turyshev:2008dr} (such as LATOR and BEACON, see Sec.~\ref{sec:mod-grav-tests}).

In this special case, when only two PPN parameters ($\gamma$, $\beta$) are considered, these parameters have a clear physical meaning. The parameter $\gamma$  represents the measure of the curvature of the space-time created by a unit rest mass; parameter  $\beta$ represents a measure of the non-linearity of the law of superposition of the gravitational fields in the theory of gravity. General relativity, when analyzed in standard PPN gauge, gives $\gamma=\beta=1$, and  the other eight parameters vanish; the theory is thus embedded in a two-dimensional space of theories. 

The Brans--Dicke theory \cite{Brans-Dicke-1961} in addition to the metric tensor, contains a scalar field and an arbitrary coupling constant $\omega$, which yields the two PPN parameter values, $\beta=1$, $\gamma= ( 1 + \omega ) / ( 2 + \omega )$, where $\omega$ is an unknown dimensionless parameter of this theory. Other general scalar-tensor theories yield different values of $\beta$  \cite{Damour_Nordtvedt_93a,Damour_Nordtvedt_93b,Turyshev:1996}.

Note that in the complete PPN framework, a particular metric theory of gravity in the PPN formalism with a specific coordinate gauge is fully characterized by means of 10 PPN parameters \cite{Turyshev:1996,Will_book93}. Thus, in addition to the parameters $\gamma$ and $\beta$, there are eight other parameters $\alpha_1, \alpha_2, \alpha_3, \zeta, \zeta_1,\zeta_2,\zeta_3$ and $\zeta_4$ (not included in Eqs.~(\ref{eq:metric}), see \cite{Will_book93} for details). The formalism uniquely prescribes the values of these parameters for the particular theory under study. Gravity experiments can be analyzed in terms of the PPN metric, and an ensemble of experiments determine the unique value for these parameters, (and hence the metric field itself). 

To analyze the motion of an $N$-body system one derives the Lagrangian function $L_{\rm N}$ \cite{Einstein-Infeld-Hoffmann-1938,Turyshev:1996,Will_book93}. Within the accuracy sufficient for most of the gravitational experiments in the solar system, this function for the motion of an $N$-body system  can be presented $L_{\rm N}$ in the following form 
\cite{Landau-Lifshitz-1988,Moyer-2003,Turyshev:2008dr}:
{}
\begin{eqnarray}
L_{\rm N}&=&\sum_i m_ic^2\Big(1-{{\dot r}_i^2\over 2c^2}
-{{\dot r}_i^4\over 8c^4}\Big)- 
\frac{1}{2}\sum_{i\not=j}{Gm_i m_j\over r_{ij}}\Bigg(1+
\frac{1+2\gamma}{2c^2}({{\dot r}_i^2}+{{\dot r}_j^2}) +
\frac{3+4\gamma}{2c^2}(\dot{\bf r}_i \dot{\bf r}_j) - 
{1\over 2c^2} ({\bf n}_{ij} \dot{\bf r}_i) ({\bf n}_{ij} \dot{\bf r}_j)\Bigg)
+ 
\nonumber\\
&+&
(\beta-{1\over 2}) \sum_{i\not=j\not=k} {G^2m_im_jm_k\over r_{ij}r_{ik}c^2}
+ {\cal O}(c^{-4}).
\label{eq:4-lagrangian} 
\end{eqnarray}

The Lagrangian in Eq.~(\ref{eq:4-lagrangian}), leads to the point-mass Newtonian and relativistic perturbative accelerations in the solar system's barycentric frame\footnote{When describing the motion of spacecraft in the solar system the models also include forces from asteroids and planetary satellites \cite{Standish_Williams_2008}.}:

\begin{eqnarray}
\ddot{\bf r}_i&=&\sum_{j\not=i}\frac{\mu_j({\bf r}_j-{\bf r}_i)}{r_{ij}^3}\bigg\{
1-\frac{2(\beta+\gamma)}{c^2}\sum_{l\not=i}\frac{\mu_l}{r_{il}}-
\frac{2\beta-1}{c^2}\sum_{k\not=j}\frac{\mu_k}{r_{jk}}+
\gamma\big(\frac{{\dot r}_i}{c}\big)^2+(1+\gamma)\big(\frac{{\dot r}_j}{c}\big)^2-\frac{2(1+\gamma)}{c^2} \dot{\bf r}_i \dot{\bf r}_j\nonumber\\
&&-~\frac{3}{2c^2}\bigg[\frac{({\bf r}_i-{\bf r}_j){\dot{\bf r}}_j}{r_{ij}}\bigg]^2+\frac{1}{2c^2}({\bf r}_j-{\bf r}_i){\ddot{\bf r}}_j\bigg\}+
\frac{1}{c^2}\sum_{j\not=i}\frac{\mu_j}{r_{ij}^3}
\Big\{\Big[{\bf r}_i-{\bf r}_j\Big]\cdot\Big[ (2+2\gamma){\dot {\bf r}}_i-(1+2\gamma){\dot {\bf r}}_j\Big]\Big\}({\dot{\bf r}}_i-{\dot{\bf r}}_j)
\nonumber\\
&&+~\frac{3+4\gamma}{2c^2}\sum_{j\not=i}\frac{\mu_j{\ddot {\bf r}}_j}{r_{ij}}+{\cal O}(c^{-4}).
\label{eq:4-26-mod}
\end{eqnarray}

To determine the orbits of the planets and spacecraft one must also describe propagation of electro-magnetic signals between any of the two points in space.  The corresponding light-time equation can be derived from the metric tensor Eq.~(\ref{eq:metric}) as below 
\begin{equation}
\label{eq:light-time}
t_2-t_1= \frac{r_{12}}{c}+(1+\gamma)\sum_i \frac{\mu_i}{c^3}
\ln\left[\frac{r_1^i+r_2^i+r_{12}^i+\frac{(1+\gamma)\mu_i}{c^2}}{r_1^i+r_2^i-r_{12}^i+\frac{(1+\gamma)\mu_i}{c^2}}\right]+{\cal O}(c^{-5}),
\end{equation}
\noindent  where $t_1$ refers to the signal transmission time, and $t_2$ refers to the reception time.  $r_{1,2}$ are the barycentric positions of the transmitter and receiver, and $r_{12}$ is their spatial separation. The terms proportional to $\propto \mu_i^2$ are important only for the Sun and are negligible for all other bodies in the solar system. 

This PPN expansion serves as a useful framework to test relativistic gravitation in the context of the gravitational experiments. The main properties of the PPN metric tensor given by Eqs.~(\ref{eq:metric}) are well established and are widely used in modern astronomical practice
\cite{Moyer-1981-1,Moyer-1981-2,Turyshev:1996,Brumberg-book-1972,Brumberg-book-1991,Standish_etal_92,Will_book93}. For practical purposes one uses the metric to derive the Lagrangian function of an $N$-body gravitating system \cite{Turyshev:1996,Will_book93} which is then used to derive the equations of motion for gravitating bodies and light, namely Eqs.~(\ref{eq:4-26-mod}) and (\ref{eq:light-time}). The general relativistic equations of motion Eq.~(\ref{eq:4-26-mod}) are then used to produce numerical codes for the purposes of construction solar system's ephemerides, spacecraft orbit determination \cite{Moyer-2003,Standish_etal_92,Turyshev:1996}, and analysis of the gravitational experiments in the solar system \cite{Turyshev:2008dr,Will_book93,Turyshev_etal_acfc_2003}. 

\subsection{PPN-Renormalized Extension of General Relativity}

Given the phenomenological success of  general relativity, it is convenient to use this theory to describe experiments. In this sense, any possible deviation from general relativity would appear as a small perturbation to this general relativistic background. Such perturbations are proportional to re-normalized PPN parameters (i.e., $\bar{\gamma}\equiv\gamma-1,\bar{\beta}\equiv\beta-1$, etc.), which are zero in general relativity, but which may have non-zero values for some gravitational theories. In terms of the metric tensor, this PPN-perturbative procedure may be conceptually presented as 
{}
\begin{equation}
\label{eq:metric2}
g_{mn}=g_{mn}^{\rm GR}+\delta g^{\rm PPN}_{mn},  
\end{equation}
where metric $g_{mn}^{\rm GR}$ is derived from Eq.~(\ref{eq:metric}) by taking the general relativistic values of the PPN parameters and where $\delta g^{\rm PPN}_{mn}$ is the PPN metric perturbation. The PPN-renormalized metric perturbation $\delta g^{\rm PPN}_{mn}$ for a system of $N$ point-like gravitational sources in four dimensions may be given as 
\begin{eqnarray}
\label{eq:metric-PPN}
\delta g^{\rm PPN}_{00}&=&-
\frac{2{\bar \gamma}}{c^4}\sum_{j\not=i}\frac{\mu_j{\dot r}^2_j}{r_{ij}}+\frac{2{\bar \beta}}{c^4}\Big(\Big[\sum_{j\not=i}\frac{\mu_j}{r_{ij}}\Big]^2+2\sum_{j\not=i}\frac{\mu_j}{r_{ij}}\sum_{k\not=j}\frac{\mu_k}{r_{jk}}\Big)+
{\cal O}(c^{-5}),\nonumber\\
\delta g^{\rm PPN}_{0\alpha}&=& \frac{2{\bar \gamma}}{c^3}\sum_{j\not=i}\frac{\mu_j{\dot {\bf r}}^\alpha_j}{r_{ij}}+
{\cal O}(c^{-5}),\\\nonumber
\delta g^{\rm PPN}_{\alpha\beta}&=&-\delta_{\alpha\beta}\frac{2{\bar \gamma}}{c^2}\sum_{j\not=i}\frac{\mu_j}{r_{ij}}
+{\cal O}(c^{-5}).  
\end{eqnarray}
\noindent Given the smallness of the current values for the PPN parameters $\bar{\gamma}$ and $\bar{\beta}$, the PPN metric perturbation $\delta g^{\rm PPN}_{mn}$ represents a very small deformation of the general relativistic background $g_{mn}^{\rm GR}$.  The expressions in Eqs.~(\ref{eq:metric-PPN}) embody the ``spirit'' of many gravitational tests assuming that general relativity provides the correct description of the experimental situation and enables the searches for small non-Einsteinian deviations.  

Similarly, one can derive the PPN-renormalized version of the Lagrangian in Eq.~(\ref{eq:4-lagrangian}):
\begin{equation}
\label{eq:largangian-PPN}
L_{\rm N}=L_{\rm N}^{\rm GR}+\delta L^{\rm PPN}_{\rm N},  
\end{equation}
where  $L_{\rm N}^{\rm GR}$ is Eq.~(\ref{eq:4-lagrangian}) taken with general relativistic values of the PPN parameters and $\delta L^{\rm PPN}_{\rm N}$ is 
{}
\begin{eqnarray}
\delta L_{\rm N}^{\rm PPN}&=&-\frac{1}{2}\frac{{\bar \gamma}}{c^2}\sum_{i\not=j}{Gm_i m_j\over r_{ij}}(\dot{\bf r}_i +\dot{\bf r}_j)^2+ 
\frac{{\bar \beta}}{c^2} \sum_{i\not=j\not=k} {G^2m_im_jm_k\over r_{ij}r_{ik}}
+ {\cal O}(c^{-4}). 
\label{eq:4-lagrangian-PPN} 
\end{eqnarray}

The equations of motion from Eq.~(\ref{eq:4-26-mod}) may also be presented in the PPN-renormalized form with explicit dependence on the PPN-perturbative acceleration terms:
\begin{equation}
\label{eq:eq-m2}
\ddot{\bf r}_i=\ddot{\bf r}_i^{\rm GR}+\delta\ddot{\bf r}_i^{\rm PPN},  
\end{equation}
where  $\ddot{\bf r}_i^{\rm GR}$ are the equations of motion from Eq.~(\ref{eq:4-26-mod}) with the values of the PPN parameters $\gamma$ and $\beta$ set to their general relativistic values of unity. Then the PPN-perturbative acceleration term $\delta\ddot{\bf r}_i^{\rm PPN}$ is given as
{}
\begin{eqnarray}
\delta\ddot{\bf r}_i^{\rm PPN}&=&\sum_{j\not=i}\frac{\mu_j({\bf r}_j-{\bf r}_i)}{r_{ij}^3}\bigg\{
 \Big(\left[\frac{m_G}{m_I}\right]_i-1\Big)
+\frac{{\dot G}}{G}\cdot (t-t_0)-
\frac{2({\bar \beta}+{\bar\gamma})}{c^2}\sum_{l\not=i}\frac{\mu_l}{r_{il}}-
\frac{2{\bar\beta}}{c^2}\sum_{k\not=j}\frac{\mu_k}{r_{jk}}+\frac{{\bar\gamma}}{c^2} (\dot{\bf r}_i-\dot{\bf r}_j)^2\bigg\}\nonumber\\
&+&\frac{2{\bar\gamma}}{c^2}\sum_{j\not=i}\frac{\mu_j}{r_{ij}^3}
\Big\{\big({\bf r}_i-{\bf r}_j\big)\cdot\big( {\dot {\bf r}}_i-{\dot {\bf r}}_j\big)\Big\}({\dot{\bf r}}_i-{\dot{\bf r}}_j)+
\frac{2{\bar\gamma}}{c^2}\sum_{j\not=i}\frac{\mu_j{\ddot {\bf r}}_j}{r_{ij}}+{\cal O}(c^{-4}).
\label{eq:4-26-mod-PPN}
\end{eqnarray}
\noindent Eq.~(\ref{eq:4-26-mod-PPN}) provides a useful framework for gravitational research. Thus, besides the terms with PPN-renormalized parameters $\bar\gamma$ and $\bar\beta$, it also contains $([{m_G}/{m_I}]_i-1)$, the parameter that signifies a possible inequality between the gravitational and inertial masses and that is needed to facilitate investigation of a possible violation of the EP (see Sec.~\ref{sec:sep}). In addition, Eq.~(\ref{eq:4-26-mod-PPN}) also includes parameter ${\dot G}/{G}$, which is needed to investigate possible temporal variation in the gravitational constant (see Sec.~\ref{sec:variation-G}). Note that in general relativity $\delta\ddot{\bf r}_i^{\rm PPN}\equiv0$.

Finally, we present the similar expressions for the light-time equation Eq.~(\ref{eq:light-time}). This equation that can be written as  $t_2-t_1=\Delta t_{12}=\Delta t_{12}^{\rm GR}+\delta\Delta t_{12}^{\rm PPN}$ with the PPN perturbation given as below
{}
\begin{equation}
\label{eq:light-time-PPN}
\delta\Delta t_{12}^{\rm PPN}= \frac{\bar\gamma}{c^3}\sum_i  \mu_i 
\ln\left[\frac{r_1^i+r_2^i+r_{12}^i+\frac{2\mu_i}{c^2}}{r_1^i+r_2^i-r_{12}^i+\frac{2\mu_i}{c^2}}\right]+{\cal O}(c^{-5}).
\end{equation}

Eqs.~(\ref{eq:eq-m2}), (\ref{eq:4-26-mod-PPN}) and (\ref{eq:light-time-PPN}) are used to focus the science objectives and to describe gravitational experiments (especially those to be conducted in the solar system) that well be discussed below. So far, general theory of relativity survived every test \cite{Turyshev-etal-2007}, yielding the ever improving values for the PPN parameters $(\gamma,\beta)$, namely $\bar\gamma=(2.1\pm2.3)\times10^{-5}$ using the data from the Cassini spacecraft taken during solar conjunction experiment \cite{Bertotti-Iess-Tortora-2003} and $\bar\beta=(1.2\pm1.1)\times10^{-4}$, which resulted from the recent analysis of the LLR data \cite{Williams-etal-2004} (see Fig.~\ref{fig:gamma-beta}).

\section{Search for New Physics Beyond General Relativity}
\label{sec:beyond}

The fundamental physical laws of Nature, as we know them today, are described by the Standard Model of particles and fields and the general theory of relativity. The Standard Model specifies the families of fermions (i.e., leptons and quarks) and their interactions by vector fields that transmit the strong, electromagnetic, and weak forces. General relativity is a tensor-field theory of gravity with universal coupling to the particles and fields of the Standard Model.  

However, despite the beauty and simplicity of general relativity and the success of the Standard Model, our present understanding of the fundamental laws of physics has several shortcomings. Although recent  progress in string theory \cite{Witten:2001,Witten:2003} is very encouraging,  the search for a realistic theory of quantum gravity remains a challenge. This continued inability to merge gravity with quantum mechanics indicates that the pure tensor gravity of general relativity needs modification or augmentation. 
The recent remarkable progress in observational cosmology has subjected the general theory of relativity to increased scrutiny by suggesting a non-Einsteinian scenario of the Universe's evolution. Researchers now believed that new physics is needed to resolve these issues.

Theoretical models of the kinds of new physics that can solve the problems described above typically involve new interactions, some of which could manifest themselves as violations of the EP, variation of fundamental constants, modification of the inverse-square law of gravity at short distances, Lorenz symmetry breaking, or large-scale gravitational phenomena.   Each of these manifestations offers an opportunity for space-based experimentation and, hopefully, a major discovery.

In this Section I present motivations for the new generation of gravitational experiments that are expected to advance the relativistic gravity research up to five orders of magnitude below the level that is currently tested by experiments \cite{Turyshev-LATOR:2004,Turyshev-etal-2006,Turyshev-etal-2007}. Specifically, I discuss theoretical models that predict non-Einsteinian behavior which can be investigated in the experiments conducted in the solar system. Such an interesting behavior led to a number of space-based experiments proposed recently to investigate the corresponding effects (see Sec.~\ref{sec:future-tests} for details). 

%
\subsection{String/M-Theory and Tensor-Scalar Extensions of General Relativity}
\label{sec:SMT}

An understanding of gravity at the quantum level will allow us to ascertain whether the gravitational ``constant'' is a running coupling constant like those of other fundamental interactions of Nature. String/M-theory \cite{Green-Schwarz-Witten-1987} hints at a negative answer to this question, given the non-renormalization theorems of supersymmetry, a symmetry at the core of the underlying principle of string/M-theory and brane models, \cite{Polchinski-95,Horava:1996ma,Horava:1995qa,Lukas-1999,Randall-Sundrum-1999}. One-loop higher--derivative quantum gravity models may permit a running gravitational coupling, as these models are asymptotically free -- a striking property \cite{Julve-1978,Fradkin-Tseytlin-1987,Avramidi-Barvinsky-1982}. In the absence of a screening mechanism for gravity, asymptotic freedom may imply that quantum gravitational corrections take effect on
macroscopic and even cosmological scales, which has some bearing on the dark matter problem \cite{Goldman-etal-1992} and, in particular, on the subject of the large scale structure of the Universe. Either way, it seems plausible to assume that quantum gravity effects manifest themselves only on cosmological scales.

Both consistency between a quantum description of matter and a geometric description of space-time, and the appearance of singularities involving minute curvature length scales indicate that a full theory of quantum gravity is needed for an adequate description of the interior of black holes and time evolution close to the Big Bang: a theory in which gravity and the associated geometry of space-time are described in the language of quantum theory. Despite major efforts in this direction, no complete and consistent theory of quantum gravity is currently available; there are, however, a number of promising candidates.

String theory is viewed as the most promising means of making general relativity compatible with quantum mechanics \cite{Green-Schwarz-Witten-1987}. The closed-string theory has a spectrum that contains as zero-mass eigenstates the graviton $g_{MN}$, the dilaton $\Phi$, and the antisymmetric second-order tensor $B_{MN}$. There are various ways to extract the physics of our four-dimensional world, and a major difficulty lies in finding a natural mechanism that fixes the value of the dilaton field, as it does not acquire a potential at any order in string-perturbation theory. However, although the usual quantum-field theories used in elementary particle physics to describe interactions, do lead to an acceptable effective (quantum) field theory of gravity at low energies, they result in models devoid of all predictive power at very high energies.

Damour \& Polyakov \cite{Damour_Polyakov_94a,Damour_Polyakov_94b} have studied a possible a mechanism to circumvent the above-mentioned difficulty by suggesting string loop-contributions, which are counted by dilaton interactions instead of by a potential. They proposed the least coupling principle (LCP), realized via a cosmological attractor mechanism (CAM) (see e.g., Refs.~\cite{Damour_Polyakov_94a,Damour_Polyakov_94b,Damour_Piazza_Veneziano_02a,Damour_Piazza_Veneziano_02b}), which can reconcile the existence of a massless scalar field in the low energy world with existing tests of general relativity (and with cosmological inflation). However, it is not yet known whether this mechanism can be realized in string theory. The authors assumed the existence of a massless scalar field $\Psi$ (i.e., of a flat direction in the potential), with gravitational strength coupling to matter. A priori, this appears phenomenologically forbidden; however, the CAM tends to drive $\Psi$ towards a value where its coupling to matter becomes naturally $\ll 1$. After dropping the antisymmetric second-order tensor and introducing fermions $\hat \psi$ and Yang--Mills fields $\hat A^{\mu}$, with field strength $\hat F_{\mu \nu}$, in a spacetime described by the metric $\hat g_{\mu \nu}$, the relevant effective low-energy four-dimensional action in the string frame can be written in the generic form as
\begin{eqnarray} 
S_{\rm eff} &=& \int d^4 x \sqrt{-\hat g} \Big\{B(\Phi)
\Big[{1 \over \alpha^{\prime}} \Big(\hat R + 4 \hat \nabla_{m} \hat \nabla^{m}\Phi - 4 (\hat \nabla \Phi)^2\Big) - 
{k \over 4} \hat F_{mn} \hat F^{mn} - \overline{\hat \psi} \gamma^{m} \hat D_{m} \hat
\psi -\frac{1}{2}(\hat\nabla\hat\chi)^2\Big]-\frac{m_\psi}{2}\chi^2\Big\}, \label{eq:string-eff-act} 
\end{eqnarray}
\noindent where\footnote{In the general case, one expects that each mater field would have a different coupling function, e.g., $B_i\rightarrow B_\Phi\not=B_F\not=B_\psi\not=B_\chi$, etc.} $B(\Phi) = e^{-2 \Phi} + c_0 + c_1 e^{2 \Phi} + c_2 e^{4 \Phi} + ..., $, $\alpha^{\prime}$ is the inverse of the string tension, $k$ is a gauge group constant, $\chi$ is the inflation field and the constants $c_0$, $c_1$, ..., can, in principle, be determined via computation. 

To recover Einsteinian gravity, one performs a conformal transformation with $g_{mn} = B(\Phi) \hat g_{mn}$, that leads to an effective action where the coupling constants and masses are functions of the rescaled dilaton, $\varphi$:
{}
\begin{eqnarray} 
S_{\rm eff} &=& \int d^4 x \sqrt{-g} \Big[{\tilde m^2_p\over 4} R - {\tilde m^2_p\over 2} (\nabla \varphi)^2 -  {\tilde m^2_p \over 2} F(\varphi) (\nabla\psi)^2-\frac{1}{2}\tilde m^2_\varphi(\chi)\chi^2
 +{k \over 4} B_F(\varphi) F_{\mu \nu} F^{\mu \nu}
+V_{\rm vac}+ ... \Big].
\label{eq:3.4} 
\end{eqnarray}

\noindent It follows that ${\tilde m^{-2}_p} = 4\pi G = {1 \over 4} \alpha^{\prime}$ and the coupling constants and masses are now dilaton-dependent, through $ g^{-2} = k B_F(\varphi)$ and $m_A = m_A[B_F(\varphi)]$.

The CAM leads to some generic predictions even without the knowledge of the specific structure of the various coupling functions, namely $m_\psi(\varphi), m_A[B_F(\varphi)], ...$. The basic assumption is that the string-loop corrections are such that there exists a {\it minimum} in (some of) the functions $m(\varphi)$ at some (finite or infinite) value $\varphi_m$. During inflation, the dynamics is governed by a set of coupled differential equations for the scale factors $\psi$ and $\varphi$. In particular, the equation of motion for $\varphi$ contains a term $\propto-\frac{\partial}{\partial \varphi} m_\chi^2(\varphi)\chi^2$. At this state of evolution (i.e., during inflation, when $\psi$ has a large vacuum expectation value) this coupling drives $\varphi$ towards the special point $\varphi_m$ where $m_\psi(\varphi)$ reaches a minimum. Once $\varphi$ has been attracted near $\varphi_m$, $\varphi$ essentially (classically) decouples from $\psi$ so that inflation proceeds as if $\varphi$ were not there. A similar attractor mechanism exists during the other phases of cosmological evolution, and tends to decouple $\varphi$ from the dominant cosmological matter. For this mechanism to efficiently decouple $\varphi$ from all types of matter, there must be a special point $\varphi_m$ to approximately minimize all the important coupling functions. A way of having such a special point in field space is to assume that $\varphi_m = +\infty$ is a limiting point where all coupling functions have finite limits. This leads to the so-called {\it runaway dilaton} scenario, in which the mere assumption that $B_i (\Phi)\simeq c^i+{\cal O}(e^{-2\Phi})$ as $\Phi \rightarrow+\infty$ implies that $\varphi_m = +\infty$ is an attractor where all couplings vanish.

This mechanism also predicts (approximately composition-independent) values for the post-Einstein parameters $\bar\gamma$ and $\bar\beta$, which parametrize  deviations from general relativity.  For simplicity, I discuss only the theories for which $g(\varphi) = q_i(\varphi) = 1$. Hence, for a theory for which the $V (\varphi)$ can be locally neglected, given that its mass is small enought that it acts cosmologically, it has been shown that in the PPN limit, that if one writes 
\begin{equation}
\ln A(\varphi) = \alpha_0(\varphi - \varphi_0) +\frac{1}{2}
\beta_0(\varphi - \varphi_0)^2 + {\cal O}(\varphi - \varphi_0)^3,
\end{equation}
where $A(\varphi)$ is the coupling function to matter and the factor that allows one to write the theory in the Einstein frame in this model $g_{mn} = A^2(\varphi) \hat g_{mn}$. Then, the two post-Einstein parameters are of the form
\begin{equation}
\bar\gamma=-\frac{2\alpha_{\rm had}^2}{1+\alpha_{\rm had}^2}\simeq -2\alpha_{\rm had}^2 
\qquad {\rm and} \qquad
\bar\beta=\frac{1}{2}\frac{\alpha_{\rm had}^2\frac{\partial\alpha_{\rm had}}{\partial\varphi}}{(1+\alpha_{\rm had}^2)^2}\simeq \frac{1}{2}\alpha_{\rm had}^2\frac{\partial\alpha_{\rm had}}{\partial\varphi}.
\end{equation}
where $\alpha_{\rm had}$ is the dilaton coupling to hadronic matter. In this model, all tests of general relativity are violated. However, all these violations are correlated. For instance, the following link between EP violations and solar-system deviations is established
\begin{equation}
\frac{\Delta a}{a}\simeq 2.6\times10^{-5}\,\bar\gamma
\end{equation}
Given that present tests of the EP place a limit on the ratio ${\Delta a}/{a}$ of the order of $10^{-13}$ (see Sec.~\ref{sec:eep}), one finds $\bar\gamma\leq6\times 10^{-9}$. Note that the upper limit given on $\bar\gamma$ by the Cassini experiment was $10^{-5}$, so in this case the necessary sensitivity has not yet been reached to test the CAM.

It is also possible that the dynamics of the quintessence field evolves to a point of minimal coupling to matter. In Ref.~\cite{Damour_Polyakov_94b} the authors showed that  $\varphi$ could be attracted towards a value $\varphi_{m}(x)$ during the matter dominated era that decoupled the dilaton from matter. For universal coupling, $f(\varphi)=g(\varphi)=q_{i}(\varphi)$ (see Eq.~(\ref{eq:sc-tensor})), this would motivate improvements in the
accuracy of the EP and other tests of general relativity. The authors of Ref.~\cite{Veneziano:2001ah} suggested that with a large number of non-self-interacting matter species, the coupling constants are determined by the quantum corrections of the matter species, and
$\varphi$ would evolve as a run-away dilaton with asymptotic value $\varphi_{m}\rightarrow\infty$. Due to the LCP, the dependence of the masses on the dilaton implies that particles fall differently in a
gravitational field, and hence are in violation of the weak form of the EP (WEP). Although, the effect (on the order of $\Delta a / a \simeq 10^{-18}$) is rather small in the solar system conditions, application of already available technology can potentially test predictions that represent a distinct experimental signature of string/M-theory. 

\begin{wrapfigure}{R}{0.42\textwidth}
  \vspace{-00pt}
  \begin{center}
    \includegraphics[width=0.42\textwidth]{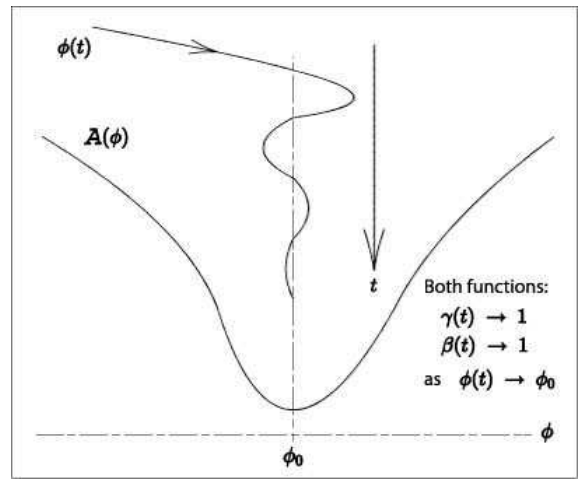}
  \end{center}
  \vspace{0pt}
\caption{\label{fig:attract}
Typical cosmological dynamics of a background scalar field is shown in the case where that field's coupling function to matter, $V(\phi)$, has an attracting point, $\phi_0$. The strength of the scalar interaction's coupling to matter is proportional to the derivative (slope) of the coupling function, so it weakens as the attracting point is approached. The Eddington parameters $\gamma$ and $\beta$ (and all higher structure parameters as well)  approach their pure tensor gravity values in this limit \cite{Damour-EspFarese-1996-2,Damour_Nordtvedt_93b,Damour_Piazza_Veneziano_02b}.  However, a small residual scalar gravity should remain because this dynamical process is not complete \cite{Turyshev-etal-2006}.
}
  \vspace{-10pt}
\end{wrapfigure}

These recent theoretical findings suggest that the present agreement between general relativity and experiment may be naturally compatible with the existence of a scalar contribution to gravity. In particular, Damour \& Nordtvedt \cite{Damour_Nordtvedt_93a,Damour_Nordtvedt_93b} (see also \cite{Damour_Polyakov_94a,Damour_Polyakov_94b} for non-metric versions of this mechanism and \cite{Damour_Piazza_Veneziano_02a,Damour_Piazza_Veneziano_02b} for a recent summary of the runaway-dilaton scenario) found that a scalar-tensor theory of gravity may contain a built-in CAM toward general relativity.  Scenarios considered by these authors assume that the scalar coupling parameter $\frac{1}{2}\bar\gamma$ was of order one in the early universe (i.e., before inflation), and show that this parameter then evolves to be close to (but not exactly equal to) zero at the present time. Fig.~\ref{fig:attract} illustrates this mechanism in greater details. 

The Eddington parameter $\gamma$, whose value in general relativity is unity, is perhaps the most fundamental PPN parameter, in that $\frac{1}{2}\bar\gamma$ is a measure of the fractional strength of the scalar gravity interaction in scalar-tensor theories of gravity \cite{Damour-EspFarese-1996-1,Damour-EspFarese-1996-2}.  Within perturbation theory for such theories, all other PPN parameters to all relativistic orders collapse to their general relativistic values in proportion to $\frac{1}{2}\bar\gamma$. Under some assumptions (see, e.g. \cite{Damour_Nordtvedt_93b}) one can estimate the likely order of magnitude of the left-over coupling strength at the present time which; this value, depending on the total mass density of the universe, can be given as $\bar\gamma \sim 7.3 \times 10^{-7}(H_0/\Omega_0^3)^{1/2}$, where $\Omega_0$ is the ratio of the current density to the closure density and where $H_0$ is the Hubble constant in units of 100 km/sec/Mpc. Compared to the cosmological constant, these scalar-field models are consistent with the supernovae observations for a lower matter density, $\Omega_0\sim 0.2$, and a higher age, $(H_0 t_0) \approx 1$. If this is indeed the case, the level $\bar\gamma \sim 10^{-6}-10^{-7}$ would be the lower bound for the present value of PPN parameter $\bar\gamma$ \cite{Damour_Nordtvedt_93a,Damour_Nordtvedt_93b}. 

Recently, Damour et al. \cite{Damour_Piazza_Veneziano_02a,Damour_Piazza_Veneziano_02b} have estimated $\frac{1}{2}\bar\gamma$ within the framework compatible with string theory and modern cosmology, confirming the results of Refs.~\cite{Damour_Nordtvedt_93a,Damour_Nordtvedt_93b}. This recent analysis discusses a scenario wherein a composition-independent coupling of a dilaton to hadronic matter produces detectable deviations from general relativity in high-accuracy light-deflection experiments in the solar system. This work assumes only some general property of the coupling functions (for large values of the field, i.e. for an ``attractor at infinity'') and then assumes that $\bar\gamma$ is on the order of one at the beginning of the controllably classical part of inflation.
Damour et al. \cite{Damour_Piazza_Veneziano_02b} showed that one can relate the present value of $\frac{1}{2}\bar\gamma$ to the cosmological density fluctuations. For the simplest inflationary potentials (favored by Wilkinson Microwave Anisotropy Probe (WMAP) mission, i.e. $m^2 \chi^2$ \cite{Bennett:2003bz}), these authors found that the present value of $\bar\gamma$ could be just below $10^{-7}$. In particular, within this framework $\frac{1}{2}(1-\gamma)\simeq\alpha^2_{\rm had}$, where $\alpha_{\rm had}$ is the dilaton coupling to hadronic matter. Its value depends on the model taken for the inflation potential $V(\chi)\propto\chi^n$, with $\chi$ again being the inflation field; the level of the expected deviations from general relativity is $\sim0.5\times10^{-7}$  for $n = 2$ \cite{Damour_Piazza_Veneziano_02b}. These predictions are based on the work in scalar-tensor extensions of gravity that are consistent with, and indeed often part of, present cosmological models.

For runaway-dilaton scenario, comparison with the minimally coupled scalar-field action,
{}
\begin{equation} S_{\phi} = {c^3\over 4\pi G}\int d^{4}x\sqrt{-g}
\left[\frac{1}{4}R +\frac{1}{2}\partial_{\mu} \phi\partial^{\mu}
\phi-V(\phi)\right],\end{equation}

\noindent reveals that the negative scalar kinetic term leads to an action equivalent to a ``ghost'' in quantum-field theory, that is referred to as ``phantom energy'' in the cosmological context \cite{Caldwell:1999ew}. Such a scalar-field model could in theory generate acceleration by the field evolving {\it up} the potential toward the maximum. Phantom fields are plagued by catastrophic ultraviolet instabilities, as particle excitations have a negative mass \cite{Cline:2003gs,Rubakov:2008nh,Sergienko:2008tf}; the fact that their energy is unbounded from below allows vacuum decay through the production of high-energy real particles as well as negative-energy ghosts that are in contradiction with the constraints on ultrahigh-energy cosmic rays \cite{Sreekumar_etal:1998}.

Such runaway behavior can potentially be avoided by the introduction of higher-order kinetic terms in the action.  One implementation of this idea is known as ``ghost condensation'' \cite{ArkaniHamed:2003uz}. In this scenario, the scalar field has a negative kinetic energy near $\dot\phi=0$, but the quantum instabilities are stabilized by the addition of higher-order corrections to the scalar field Lagrangian of the form $(\partial_{\mu} \phi\partial^{\mu} \phi)^{2}$. The ``ghost'' energy is then bounded from below, and stable evolution of the dilaton occurs with $w\ge-1$ \cite{Piazza:2004df}.  The gradient $\partial_\mu\phi$ is non-vanishing in the vacuum, violating Lorentz invariance; this may have important consequences in cosmology and in laboratory experiments.

The analyses discussed above predict very small (ranging from $10^{-5}$ to $5\times10^{-8}$ for $\frac{1}{2}\bar\gamma$) observable post-Newtonian deviations from general relativity in the solar system, thereby motivating new generation of advanced gravity experiments. In many cases, such tests would require reaching the accuracy needed to measure effects of the next post-Newtonian order ($\propto G^2$) \cite{Turyshev-etal-2007,Turyshev-LATOR:2003}, promising important outcomes for the twenty-first century fundamental physics. 

\subsection{Observational Motivations for Higher Accuracy Tests of Gravity}
\label{sec:mot_experiment}

Recent astrophysical measurements of the angular structure of the cosmic microwave background (CMB) \cite{deBernardis_CMB2000}, the masses of large-scale structures \cite{Peacock_LargeScale01}, and the luminosity distances of type Ia supernovae \cite{Perlmutter:1998np,Riess_supernovae98} have placed stringent constraints on the cosmological constant $\Lambda$ and  have also led to a revolutionary conclusion: The expansion of the universe is accelerating. The implication of these observations for cosmological models is that a classically evolving scalar field currently dominates the energy density of the universe. Such models have been shown to share the advantages of  $\Lambda$, namely (a) compatibility with the spatial flatness predicted inflation, (b) a universe older than the standard Einstein-de Sitter model, and, (c) combined with cold dark matter (CDM), predictions for large-scale structure formation in good agreement with data from galaxy surveys. As well as imprinting their distinctive signature on the CMB anisotropy, scalar-field models remain viable and should be testable in the near future. This completely unexpected discovery demonstrates the importance of testing important ideas about the nature of gravity. We are presently in the ``discovery'' phase of this new physics, and although there are many theoretical conjectures as to the origin of a non-zero $\Lambda$, it is essential that we exploit every available opportunity to elucidate the physics at the root of the observed phenomena.

Description of quantum matter in a classical gravitational background poses interesting challenges, notably the possibility that the zero-point fluctuations of the matter fields generate a non-vanishing vacuum energy density $\rho_{\rm vac}$, that corresponds to the term $-\sqrt{-g}\rho_{\rm vac}$, in ${\cal S}_{\rm SM}$ Eq.~(\ref{eq:action-SM}) \cite{Weinberg-1989:RevModPhys}. This is equivalent to adding a ``cosmological constant'' term $+\Lambda g_{mn}$ on the left-hand side of Einstein's equations Eq.~(\ref{eq:Einstein-eq}), with $\Lambda = 8\pi G_M \rho_{\rm vac}/c^4$. Recent cosmological observations suggest a positive value of $\Lambda$ corresponding to $\rho_{\rm vac} \approx (2.3 \times 10^{-3} eV)^4$ \cite{Spergel-etal-2006}.  Such a small value has a negligible effect on the solar system dynamics and relevant gravitational tests. Quantizing the gravitational field itself poses a challenge because of the perturbative non-renormalizability of Einstein's Lagrangian \cite{Weinberg-1989:RevModPhys,Wilczek:2006}. Superstring theory offers a promising avenue toward solving this challenge.

There is now a great deal of evidence indicating that over 70\% of the critical density of the universe is in the form of a ``negative-pressure'' dark energy component; we have no understanding of its origin or nature. The fact that the expansion of the universe is currently undergoing a period of acceleration has been well tested: The expansion has been directly measured from the light curves of several hundred type Ia supernovae \cite{Perlmutter:1998np,Riess_supernovae98,Tonry:2003zg}, and has been independently inferred from observations of CMB by the WMAP satellite \cite{Bennett:2003bz} and other CMB experiments \cite{Halverson:2001yy,Netterfield:2001yq}. Cosmic speed-up can be accommodated within general relativity by invoking a mysterious cosmic fluid with large negative pressure, dubbed dark energy. The simplest possibility for dark energy is a cosmological constant; unfortunately, the smallest estimates for its value are 55 orders of magnitude too large (for reviews see \cite{Carroll:2000fy,Peebles:2002gy}). Most of the theoretical studies operate in the shadow of the cosmological constant problem, the most embarrassing hierarchy problem in physics. This fact has motivated a host of other possibilities, most of which assume $\Lambda=0$, with the dynamical dark energy being associated with a new scalar field (see \cite{Carroll:2003wy,Carroll:2004de} and references therein). However, none of these suggestions is compelling and most have serious drawbacks. Given the magnitude of this problem, a number of authors have considered the possibility that cosmic acceleration is not due to a particular substance, but rather that it arises from new gravitational physics (see discussion in \cite{Carroll:2000fy,Peebles:2002gy,Carroll:2003wy}).
In particular, certain extensions to general relativity in a low energy regime \cite{Carroll:2003wy,Capozziello:2005bu,Carroll:2004de} were shown to predict an experimentally consistent universe evolution  without the need for dark energy  \cite{Bertolami-Paramos-Turyshev-2007}. These dynamical models are expected to explain the observed acceleration of the universe without dark energy, but may produce measurable gravitational effects on the scales of the solar system.  

\subsection{Modified Gravity as an Alternative to Dark Energy}
\label{sec:de}

Certain modifications of the Einstein--Hilbert action Eq.~(\ref{eq:hilb-ens-action}) by introducing terms that diverge as the scalar curvature goes to zero could mimic dark energy \cite{Carroll:2003wy,Carroll:2004de}.  Recently, models involving inverse powers of the curvature have been proposed as an alternative to dark energy. In these models there are more propagating degrees of freedom in the gravitational sector than the two contained in the massless graviton in general relativity. The simplest models of this kind add inverse powers of the scalar curvature to the action ($\Delta {\cal L}\propto 1/R^n$), thereby introducing a new scalar excitation in the spectrum. For the values of the parameters required to explain the acceleration of the Universe this scalar field is almost massless in vacuum; this could lead to a possible conflict with the solar system experiments. 

However, models that involve inverse powers of other invariants, in particular those that diverge for $r\rightarrow 0$ in the Schwarzschild solution, generically recover an acceptable weak-field limit at short distances from sources by means of a screening or shielding of the extra degrees of freedom at short distances \cite{Navarro:2005da}. Such theories can lead to late-time acceleration, but they typically result in one of two problems: either they are in conflict with tests of general relativity in the solar system, due to the existence of additional dynamical degrees
of freedom \cite{Chiba:2003ir}, or they contain ghost-like degrees of freedom that seem difficult to reconcile with fundamental theories.   

The idea that the cosmic acceleration of the Universe may be caused by modification of gravity at very large distances, and not by a dark energy source, has recently received a great deal of attention (see \cite{Rubakov:2008nh,Sotiriou:2008rp}). Such a modification could be triggered by extra space dimensions, to which gravity spreads over cosmic distances. In addition to being testable by cosmological surveys, modified gravity predicts testable deviations in planetary motions, providing new motivations for a new generation of advanced gravitational experiments in space \cite{Turyshev-etal-2007,Turyshev:2008dr}.
An example of recent theoretical progress is  the Dvali-Gabadadze-Porrati (DGP) brane-world model, which explores the possibility that we live on a brane embedded in a large extra dimension, and where the strength of gravity in the bulk is substantially less than that on the brane \cite{Dvali-Gabadadze-Porrati-2000}. Although such a theory can lead to perfectly conventional gravity on large scales, it is also possible to choose the dynamics in such a way that new effects show up exclusively in the far infrared, thereby providing a mechanism to explain the acceleration of the universe \cite{Perlmutter:1998np,Riess_supernovae98}.  Interestingly, DGP gravity and other modifications of general relativity hold out the possibility of having interesting and testable predictions that distinguish them from models of dynamical dark energy. One outcome of this work is that the physics of the accelerating universe may be deeply tied to the properties of gravity on relatively short scales, from millimeters to astronomical units \cite{Dvali-Gabadadze-Porrati-2000,Dvali-Gruzinov-Zaldarriaga-2003}.

Although many effects predicted by gravity modification models are suppressed within the solar system, there are measurable effects induced by some long-distance modifications of gravity \cite{Dvali-Gabadadze-Porrati-2000}. For instance, in the case of the precession of the planetary perihelion in the solar system, the anomalous perihelion advance, $\Delta \phi$, induced by a small correction, $\delta U_N$, to Newton's potential, $U_N$, is given in radians per revolution \cite{Dvali-Gruzinov-Zaldarriaga-2003} by
$\Delta \phi \simeq \pi r {d \over dr} [r^2 {d \over dr}({\delta U_N \over rU_N})].$ The most reliable data regarding the planetary perihelion advances come from the inner planets of the solar system, where a majority of the corrections are negligible. However, LLR offers an interesting possibility to test for these new effects \cite{Williams-etal-2004}. Evaluating the expected magnitude of the effect to the Earth-Moon system, one predicts an anomalous shift of $\Delta \phi \sim 10^{-12}$ \cite{Dvali-Gruzinov-Zaldarriaga-2003}, compared with the achieved accuracy of $2.4\times 10^{-11}$. Therefore, the theories of gravity modification raise an intriguing possibility of discovering new physics that could be addressed with the new generation of astrometric measurements \cite{Turyshev:2008dr}.

\subsection{Scalar Field Models as Candidates for Dark Energy}
\label{sec:sc-models-de}

One of the simplest candidates for dynamical dark energy is a scalar field, $\varphi$, with an extremely low-mass and an effective potential $V(\varphi)$.  If the field is rolling slowly, its persistent potential energy is responsible for creating the late epoch of inflation we observe today. For the models that include only inverse powers of the curvature, other than the Einstein-Hilbert term, it is possible that in regions where the curvature is large the scalar has a large mass that could make the dynamics similar to those of general relativity \cite{Cembranos:2005fi}. At the same time, the scalar curvature, although larger than its mean cosmological value, is very small in the solar system, thereby satisfying constraints set by the gravitational tests performed to date \cite{Erickcek:2006vf,Nojiri:2006gh,Turyshev-dilaton-1995,Turyshev-BH-1996,Silaev-Turyshev-1997}. Nevertheless, it is not clear whether these models may be regarded as a viable alternative to dark energy.

Effective scalar fields are prevalent in supersymmetric field theories and string/M-theory. For example, string theory predicts that the vacuum expectation value of a scalar field, the dilaton, determines the relationship between the gauge and gravitational couplings. A general, low energy effective action for the massless modes of the dilaton can be cast as a scalar-tensor theory (as in Eq.~(\ref{eq:sc-tensor})) with a vanishing potential, where $f(\varphi)$, $g(\varphi)$, and $q_{i}(\varphi)$ are the dilatonic couplings to gravity, the scalar kinetic term, and the gauge and matter fields, respectively, which encode the effects of loop effects and potentially non-perturbative corrections.

A string-scale cosmological constant or exponential dilaton potential in the string frame translates into an exponential potential in the Einstein frame. Such quintessence potentials \cite{Wetterich:1988,Ratra-Peebles-1988,Wetterich:2004ss,Peebles:2002gy,Wetterich:2004pv} can have scaling \cite{Ferreira_Joyce:1997}, and tracking \cite{Zlatev_Wang_Steinhardt:1999} properties that allow the scalar field energy density to evolve alongside the other matter constituents. A problematic feature of scaling potentials \cite{Ferreira_Joyce:1997} is that they do not lead to accelerative expansion, as the energy density simply scales with that of matter. Alternatively, certain potentials can predict a dark energy density that alternately dominates the Universe and decays away; in such models, the acceleration of the Universe is transient \cite{Albrecht_Skordis:2000,Dodelson_Kaplinghat_Stewart:2000,Bento_Bertolami_Santos:2002}. Collectively, quintessence potentials predict that the density of the dark energy dynamically evolves over time, in contrast to the cosmological constant. Similar to a cosmological constant, however, the scalar field is expected to have no significant density perturbations within the causal horizon, so they contribute little to the evolution of the clustering of matter in large-scale structure \cite{Ferreira_Joyce:1998}.

In addition to couplings to ordinary matter, the quintessence field may have nontrivial couplings to dark matter \cite{Farrar_Peebles:2004,Bertolami-Paramos-Turyshev-2007}. Non-perturbative string-loop effects do not lead to universal couplings, although it is possible that the dilaton decouples more slowly from dark matter than it does from gravity and fermions. This coupling can provide a mechanism to generate acceleration with a scaling potential while also being consistent with EP tests. It can also explain why acceleration began to occur only relatively recently, by being triggered by the non-minimal coupling to the CDM, rather than by a feature in the effective potential \cite{Bean:2000zm,Gasperini:2001pc}. Such couplings can not only generate acceleration, but can also modify structure formation through the coupling to CDM density fluctuations \cite{Bean:2001ys} and adiabatic instabilities \cite{Bean:2007nx,Bean:2007ny}, in contrast to minimally coupled quintessence models. Dynamical observables that are sensitive to the evolution in matter perturbations as well as to the expansion of the Universe, such as (a) the matter power spectrum as measured by large-scale surveys and (b) weak lensing convergence spectra, could distinguish non-minimal couplings from theories with minimal effect on clustering.  

In the next section I will discuss the new effects predicted by the theories and models discussed above.  I will also present a list of experiments that were proposed to test these important predictions in dedicated space experiments.

\section{Search for a new theory of gravity with space-based experiments}
\label{sec:future-tests} 

It is well known that work on the general theory of relativity began with the EP, in which gravitational acceleration was a priori held indistinguishable from acceleration caused by mechanical forces; as a consequence, gravitational mass was therefore identical to inertial mass. Since Newton's time, the question about the equality of inertial and passive gravitational masses has risen in almost every theory of gravitation. Einstein elevated this identity, which was implicit in Newton's gravity, to a guiding principle in his attempts to explain both electromagnetic and gravitational acceleration according to the same set of physical laws \cite{Einstein-1907,Einstein-1915,Einstein-1916,Frederiks:1999,Kobzarev:1990}. Thus, almost 100 years ago Einstein postulated that not only mechanical laws of motion, but also all non-gravitational laws should behave in freely falling frames as if gravity were absent. It is this principle that predicts identical accelerations of compositionally different objects in the same gravitational field, and it also allows gravity to be viewed as a geometrical property of spacetime--leading to the general relativistic interpretation of gravitation.  

\begin{wrapfigure}{R}{0.46\textwidth}
  \vspace{-20pt}
  \begin{center}
    \includegraphics[width=0.46\textwidth]{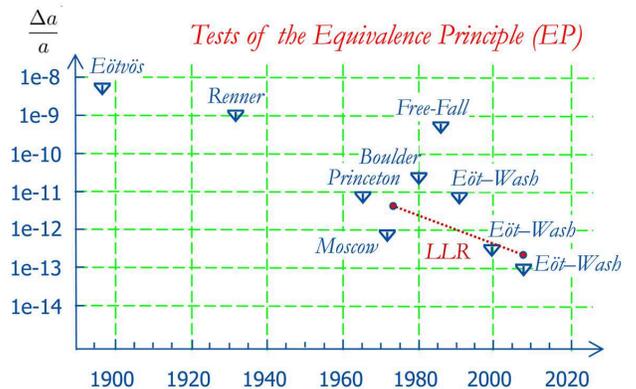}
  \end{center}
  \vspace{-10pt}
  \caption{Progress in the tests of the equivalence principle (EP) since the early twentieth century  \cite{Turyshev-etal-2007,Turyshev:2008dr}.
}
  \vspace{-5pt}
\end{wrapfigure}

Remarkably, the EP has been (and still is!) a focus of gravitational research for more than 400 years \cite{Williams-etal-2005}.  Since the time of Galileo we have known that objects of different mass and composition accelerate at identical rates in the same gravitational field.  From 1602 to 1604, through his study of inclined planes and pendulums, Galileo formulated a law of falling bodies that led to an early empirical version of the EP.  However, these famous results were not published for another 35 years.  It took an additional 50 years before a theory of gravity describing these and other early gravitational experiments was published by Newton in his Principia in 1687.  On the basis of his second law, Newton concluded that the gravitational force is proportional to the mass of the body on which it acted; from his third law, he postulated the gravitational force is proportional to the mass of its source. 

Newton was aware that the {\it inertial mass} $m_I$ in his second law, ${\bf F} = m_I {\bf a}$, might not be the same as the {\it gravitational mass} $m_G$ relating force to gravitational field  ${\bf F} = m_G {\bf g}$. Indeed, after rearranging these two equations, we find ${\bf a} = (m_G/m_I){\bf g}$ and thus, in principle, materials with different values of the ratio $m_G/m_I$ could accelerate at different rates in the same gravitational field.  Newton tested  this possibility with simple pendulums of the same length but with different masses and compositions, but he found no difference in their periods.  On this basis Newton concluded that $m_G/m_I$ was constant for all matter; and that, by a suitable choice of units, the ratio could always be set to one, i.e. $m_G/m_I = 1$. Bessel subsequently tested this ratio more accurately, and then in a definitive 1889 experiment E\"otv\"os was able to experimentally verify this equality of the inertial and gravitational masses to an accuracy of one part in $10^9$ \cite{Eotvos_1890,Eotvos_etal_1922,Bod_etal_1991}. 

Today, more than 320 years after Newton proposed a comprehensive approach to studying the relation between the two masses of a body, this relation remains to be the subject of numerous theoretical and experimental investigations.  The question regarding the equality of inertial and passive gravitational masses has arisen in almost every theory of gravitation.  In 1915 the EP became a part of the foundation of Einstein's general theory of relativity; subsequently, many experimental efforts focused on testing the EP in the search for limits of general relativity.  Thus, the early tests of the EP were further improved by Dicke and his colleagues \cite{Roll_etal_1964} to one part in $10^{11}$. Most recently, a University of Washington group \cite{Baessler_etal_1999,Adelberger_2001} improved upon Dicke's verification of the EP by several orders of magnitude, reporting $(m_G/m_I - 1) = 1.4 \times 10^{-13}$, thereby confirming Einstein's intuition.

In a 1907 paper, using the early version of the EP \cite{Einstein-1907} Einstein made important preliminary predictions regarding the influence of gravity on light propagation; these predictions presented the next important step in the development of his theory.  He realized that a ray of light coming from a distant star would appear to be attracted by solar mass while passing close to the Sun.  As a result, the ray's trajectory is bent twice as much in the direction towards the Sun compared to the same trajectory analyzed with Newton's theory (see discussion in Sec.~\ref{sec:intro}). In addition, light radiated by a star would interact with the star's gravitational potential, resulting in the radiation shifting slightly toward the infrared end of the spectrum. So the original EP, as described by Einstein, concluded that free-fall and inertial motion were physically equivalent which constitutes the weak the form of the EP (WEP). 

In about 1912, Einstein (with the help of mathematician Marcel Grossmann) began a new phase of his gravitational research by phrasing his work in terms of the tensor calculus of Tullio Levi-Civita and Gregorio Ricci-Curbastro. The tensor calculus greatly facilitated calculations in four-dimensional space-time, a notion that Einstein had obtained from Hermann Minkowski's 1907 mathematical elaboration of Einstein's own special theory of relativity. Einstein called his new theory the general theory of relativity.  After a number of false starts, he published the definitive field equations of his theory in late 1915 \cite{Einstein-1915,Einstein-1916}.  
Since that time, physicists have endeavored to understand and verify various predictions of the general theory of relativity with ever increasing accuracy (for review of various gravitational experiments available during the period of 1970-2000, see \cite{Braginsky-Rudenko:1970,Konopleva:1977,Rudenko:1978,Kobzarev:1990,Will_book93})

Note that, although the EP guided the development of general relativity, it is not a founding principle of relativity but rather a simple consequence of the geometrical nature of the theory. In general relativity, test objects in free-fall follow geodesics of spacetime, and what we perceive as the force of gravity is instead a result of our being unable to follow those geodesics of spacetime, because the mechanical resistance of matter prevents us from doing so.

Below we discuss on the space-based gravitational experiments aiming to test the various aspects of the EP, tests of Lorentz and position invatiancies, search for variability of the fundamental constants, tests of the gravitational inverse square law and tests of alternative and modified-gravity theories. 

\subsection{Tests of the Equivalence Principle}
\label{sec:eep}

Since Einstein developed general relativity, there was a need to develop a framework to test the theory against other possible theories of gravity compatible with special relativity. This was developed by Robert Dicke \cite{Dicke:1959,Dicke:1961} as part of his program to test general relativity. Two new principles were suggested, the so-called Einstein EP (EEP) and the strong EP (SEP), each of which assumes the WEP as a starting point. They only differ in whether or not they apply to gravitational experiments.

The EEP states that the result of a local non-gravitational experiment in an inertial frame of reference is independent of the velocity or location in the universe of the experiment. This is a kind of Copernican extension of Einstein's original formulation, which requires that suitable frames of reference all over the universe behave identically. It is an extension of the postulates of special relativity in that it requires that dimensionless physical values such as the fine-structure constant and electron-to-proton mass ratio be constant. From the theoretical standpoint, the EEP \cite{Damour_Nordtvedt_93a,Damour_Nordtvedt_93b,Williams-etal-2004,Williams-etal-2005} is at the foundation of the general theory of relativity; therefore, testing the principle is very important. As far as the experiment is concerned, the EEP includes three testable hypotheses: 
\begin{itemize}
\item[i)] 
Universality of free fall (UFF), which states that freely falling bodies have the same acceleration in the same gravitational field independent of their compositions (see Sec.~\ref{sec:eep}),
\item [ii)] 
Local Lorentz invariance (LLI), which suggests that clocks' rates are independent of the clock's velocities (see Sec.~\ref{sec:LLI}), and
\item [iii)]
Local position invariance (LPI), which postulates that clocks' rates are also independent of their spacetime positions (see Sec.~\ref{sec:LPI}). 
\end{itemize}
Using these three hypotheses Einstein deduced that gravity is a geometric property of spacetime \cite{Dicke:1959,Will-lrr-2006-3}. One can test the validity of both the EP and the field equations that determine the geometric structure created by a mass distribution. There are two different ``flavors'' of the EP,  the weak and the strong forms (WEP and SEP, respectively), which are being tested in various experiments performed with laboratory test masses and with bodies of astronomical sizes \cite{Williams-etal-2005}.

\subsubsection{The Weak  Equivalence Principle}
\label{sec:wep}

The \emph{weak form of the EP} (the WEP, also known as the UFF) holds that the gravitational properties of strong and electro-weak interactions obey the EP. In this case the relevant test-body differences are their fractional nuclear-binding differences, their neutron-to-proton ratios, their atomic charges, etc.  Furthermore, the equality of gravitational and inertial masses implies that different neutral massive test bodies have the same free-fall acceleration in an external gravitational field, and therefore in freely falling inertial frames the external gravitational field appears only in the form of a tidal interaction \cite{Singe-1960}. Apart from these tidal corrections, freely falling bodies behave as though external gravity were absent \cite{Anderson-etal-1996}. 

General relativity and other metric theories of gravity assume that the WEP is exact.  However, many extensions of the Standard Model that contain new macroscopic-range quantum fields predict quantum exchange forces that generically violate the WEP because they couple to generalized ``charges'' rather than to mass/energy as does gravity \cite{Damour_Nordtvedt_93a,Damour_Nordtvedt_93b,Damour_Polyakov_94a,Damour_Polyakov_94b,Damour_Piazza_Veneziano_02a,Damour_Piazza_Veneziano_02b}. 

In a laboratory, precise tests of the EP can be made by comparing the free fall accelerations, $a_1$ and $a_2$, of different test bodies. When the bodies are at the same distance from the source of the gravity, the expression for the EP takes the elegant form
\begin{equation} 
{\Delta a \over a} = {2(a_1- a_2) \over a_1 + a_2} =
\left[{m_G \over m_I}\right]_1 -\left[{m_G \over m_I}\right]_2 = \Delta\left[{m_G \over m_I}\right], 
\label{WEP_da} \end{equation}
\noindent where $m_G$ and $m_I$ are the gravitational and inertial masses of each body, respectively. 
The sensitivity of the EP test is determined by the precision of the differential acceleration measurement divided by the degree to which the test bodies differ (\textit{e.g.} composition).

Various experiments have been performed to measure the ratios of gravitational to inertial masses of bodies. Recent experiments on bodies of laboratory dimensions have verified the WEP to a fractional precision of $\Delta(m_G/m_I) \lesssim 10^{-11}$ by \cite{Roll_etal_1964}, to $\lesssim 10^{-12}$ by \cite{Braginsky-Panov:1972,Su-etal:1994} and more recently to a precision of $\lesssim 1.4\times 10^{-13}$ \cite{Adelberger_2001}. The accuracy of these experiments is high enough to confirm that the strong, weak, and electromagnetic interactions each contribute equally to the passive gravitational and inertial masses of the laboratory bodies.  

Currently, the most accurate results in testing the WEP have been reported by ground-based laboratories \cite{Williams-etal-2005,Baessler_etal_1999}. The most recent result \cite{Adelberger_2001,Schlamminger-etal-2007} for the fractional differential acceleration between beryllium and titanium test bodies was given by the E\"ot-Wash group\footnote{The E\"ot-Wash group at the University of Washington in Seattle has developed new techniques in high-precision studies of weak-field gravity and searches for possible new interactions weaker than gravity. See {\tt http://www.npl.washington.edu/eotwash/} for details.} as
$\Delta a/a=(1.0 \pm 1.4) \times 10^{-13}$.  A review of the most recent laboratory tests of gravity can be found in Ref. \cite{Gundlach-etal-2007}. Significant improvements in the tests of the EP are expected from dedicated space-based experiments \cite{Turyshev-etal-2007}.

\begin{wrapfigure}{R}{0.16\textwidth}
  \vspace{-15pt}
  \begin{center}
    \includegraphics[width=0.16\textwidth]{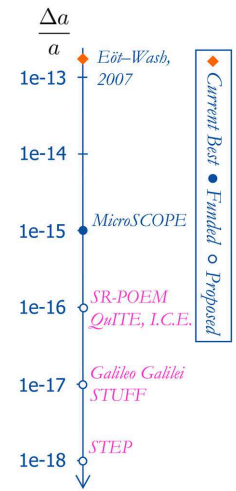}
  \end{center}
  \vspace{-10pt}
  \caption{Anticipated progress in the tests of the WEP \cite{Turyshev-etal-2007,Turyshev:2008dr}.}
  \vspace{-5pt}
\label{fig:future-WEP-tests}
\end{wrapfigure}

The composition-independence of acceleration rates of various masses toward the Earth can be tested in space-based laboratories to a precision of many additional orders of magnitude, down to levels at which some models of the unified theory of quantum  gravity, matter, and energy suggest a possible violation of the EP \cite{Damour_Nordtvedt_93a,Damour_Nordtvedt_93b,Damour_Polyakov_94a,Damour_Polyakov_94b,Damour_Piazza_Veneziano_02a,Damour_Piazza_Veneziano_02b}.  In some scalar-tensor theories, the strength of EP violations and the magnitude of the fifth force mediated by the scalar can be drastically larger in space than on the ground \cite{Mota-Barrow-2004-2}, providing further justification for space deployment.  Importantly, many of these theories predict observable violations of the EP at various levels of accuracy ranging from $10^{-13}$ to $10^{-16}$. Therefore, even a confirmation of no EP-violation will be exceptionally valuable, as it will place useful constraints on the range of possibilities in the development of a unified physical theory. 

Compared with Earth-based laboratories, experiments in space can benefit from a range of conditions, including free fall and significantly reduced contributions due to seismic, thermal, and other non-gravitational noise (see Appendix A in \cite{Turyshev-etal-2007}). As a result, many experiments have been proposed to test the EP in space. Below I present only a partial list of these missions.  Furthermore, to illustrate the use of different technologies, I discuss only the most representative concepts while not going into technical details of these experiments.  

The Micro-Satellite \`a tra\^in\'ee Compens\'ee pour l'Observation du Principe d'Equivalence (MicroSCOPE) mission\footnote{\label{foot:microscope}See \tt http://microscope.onera.fr/ for details on MicroSCOPE mission.}  is a room-temperature EP experiment in space that utilizes electrostatic differential accelerometers \cite{Touboul-Rodrigues-2001}.  The mission is currently under development by Centre National d'Etudes Spatiales (CNES)\footnote{Centre National d'Etudes Spatiales (CNES) -- the French Space Agency, see website at: {\tt http://www.cnes.fr/}} and the European Space Agency (ESA), and is scheduled for launch in 2010.  The design goal is to achieve a differential acceleration accuracy of $10^{-15}$. MicroSCOPE's electrostatic differential accelerometers are based on flight heritage designs from the CHAMP, GRACE, and GOCE missions.\footnote{Several gravity missions were recently developed by the German National Research Center for Geosciences (GFZ).  Among them are CHAMP (Gravity and Magnetic Field Mission); GRACE (Gravity Recovery And Climate Experiment Mission, together with NASA); and GOCE (Global Ocean Circulation Experiment), together with ESA and other European countries. See {\tt http://www.gfz-potsdam.de/pb1/op/index\_GRAM.html}}

The Principle of Equivalence Measurement (PO\-EM) experiment
\cite{Reasenberg-Phillips-2007} is a ground-based test of the WEP and now is under development.  It will be able to detect a violation of the EP with a fractional acceleration accuracy of 5 parts in 10$^{14}$ in a short experiment (i.e., a few days long) and with a three- to tenfold better accuracy in a longer experiment.  The experiment makes use of optical distance measurement (by tracking frequency gauge (TFG) laser gauge \cite{Phillips-Reasenberg-2005}) and  will be  advantageously sensitive to short-range forces with a characteristic length scale of $\lambda < 10$~km. SR-POEM, a POEM-based  proposed room-temperature test of the WEP during a sub-orbital flight on a sounding rocket, was also recently proposed \cite{Turyshev-etal-2007}. It is anticipated to be able to search for a violation of the EP with a single-flight  accuracy of 1 part in 10$^{16}$. Extension to higher accuracy in an orbital mission is under study. Additionally, the Space Test of Universality of Free Fall (STUFF) \cite{Turyshev-etal-2007} is a recent study of a space-based experiment that relies on optical metrology and proposes to reach an accuracy of 1 part in 10$^{17}$ in testing the EP in space.

The Quantum Interferometer Test of the Equivalence Principle (QuITE) \cite{Kasevich-Maleki-2003} is a proposed cold-atom-based test of the EP in space. QuITE intends to measure the absolute single axis differential acceleration, with an accuracy of 1 part in $10^{16}$, by utilizing two co-located matter wave interferometers with different atomic species.\footnote{Compared to the ground-based conditions, space offers a factor of nearly $10^3$ improvement in the integration times in observation of the free-falling atoms (i.e., progressing from ms to sec).  The longer integration times translate into the accuracy improvements \cite{Turyshev-etal-2007}.} QuITE will improve the current EP limits set in similar experiments conducted in ground-based laboratory conditions\footnote{Its ground-based analog, called ``Atomic Equivalence Principle Test (AEPT)'', is currently being built at Stanford University. AEPT is designed to reach sensitivity of one part in $10^{15}$.} \cite{Peters-Chung-Chu-2001,Fray-etal-2004} by nearly seven to nine orders of magnitude. Similarly, the Interf\'erom\'etrie \`a Source Coh\'erente pour Applications dans l'Espace (I.C.E.) project\footnote{Interf\'erom\'etrie \`a Source Coh\'erente pour Applications dans l'Espace (I.C.E.), see {\tt http://www.ice-space.fr}} supported by CNES in France aims to develop a high-precision accelerometer based on coherent atomic sources in space \cite{Nyman-etal-2006} with an accurate test of the EP as one of its main objectives.

The Galileo Galilei (GG) mission \cite{Nobili-etal-2007} is an Italian space experiment\footnote{Galileo Galilei (GG) website: {\tt http://eotvos.dm.unipi.it/nobili}} proposed to test the EP at room temperature with an accuracy of 1 part in $10^{17}$.  The key instrument of GG is a differential accelerometer made of weakly coupled coaxial, concentric test cylinders that spin rapidly around the symmetry axis and are sensitive in the plane perpendicular to it. GG  is included in the National Aerospace Plan of the Italian Space Agency (ASI) for implementation in the near future. 

The Satellite Test of Equivalence Principle (STEP) mission \cite{step-2001-1,step-2001-2} is a proposed test of the EP to be conducted from a free-falling platform in space provided by a drag-free spacecraft orbiting the Earth.  STEP will test the composition independence of gravitational acceleration for cryogenically controlled test masses by searching for a violation of the EP with a fractional acceleration accuracy of 1 part in 10$^{18}$.  As such, this ambitious experiment will be able to test very precisely for the presence of any new non-metric, long range physical interactions.

This impressive evidence and the future prospects of testing the WEP for laboratory bodies is incomplete for astronomical body scales. The experiments searching for WEP violations are conducted in laboratory environments that utilize test masses with negligible amounts of gravitational self-energy; therefore, a large-scale experiment is needed to test the postulated equality of gravitational self-energy contributions to the inertial and passive gravitational masses of the bodies \cite{Nordtvedt_1968a}. Once the self-gravity of the test bodies is non-negligible (which is currently true only for bodies of astronomical sizes), the corresponding experiment test the ultimate version of the EP -- the SEP.

\subsubsection{The Strong Equivalence Principle}
\label{sec:sep}

In its \emph{strong form the EP} (the SEP) is extended to cover the gravitational properties resulting from gravitational energy itself \cite{Williams-etal-2005}.  It is an assumption about the way that gravity begets gravity, i.e. about the non-linear property of gravitation. Although general relativity assumes that the SEP is exact, alternate metric theories of gravity---such as those involving scalar fields and other extensions of gravity theory---typically violate the SEP. For the SEP case, the relevant test-body differences are the fractional contributions to their masses by gravitational self-energy. Because of the extreme weakness of gravity, SEP test bodies must have astronomical sizes. 

The SEP states that the results of any local experiment, gravitational or not, in an inertial frame of reference are independent of where and when in the universe it is conducted. This is the only form of the EP that applies to self-gravitating objects (such as stars), which have substantial internal gravitational interactions. It requires that the gravitational constant be the same everywhere in the universe and is incompatible with a fifth force. It is much more restrictive than the EEP. General relativity is the only known theory of gravity compatible with this form of the EP.

Nordtvedt \cite{Nordtvedt_1968a,Ken_LLR68,Nordtvedt_1970} suggested several solar system experiments for testing the SEP. One of these was the lunar test. Another, a search for the SEP effect in the motion of the Trojan asteroids, was carried out by
\cite{Orellana_Vucetich_1993}. Interplanetary spacecraft tests have been considered by \cite{Anderson-etal-1996} and discussed \cite{AndersonWilliams01}. An experiment employing existing binary pulsar data has been proposed \cite{Damour_Schafer_1991}. It was pointed out that binary pulsars may provide an excellent possibility for testing the SEP in the new regime of strong self-gravity \cite{Damour-EspFarese-1996-2}, however the corresponding tests have yet to reach competitive accuracy \cite{Kramer-etal-2006}.

The PPN formalism \cite{Nordtvedt_1968b,Will-1971-I,Will-1971-II,Will-Nordtvedt-72-I,Will-Nordtvedt-72-II} describes the motion of celestial bodies in a theoretical framework common to a wide class of metric theories of gravity. 
To facilitate investigation of a possible violation of the SEP, Eq.~(\ref{eq:4-26-mod}) allows for a possible inequality of the gravitational and inertial masses, given by the parameter $[{m_G}/{m_I}]_i$, which in the PPN formalism is expressed \cite{Nordtvedt_1968a,Nordtvedt_1968b} as
{}
\begin{equation}
\left[\frac{m_G}{m_I}\right]_{\tt SEP} = 1 + 
\eta\Big(\frac{E}{mc^2}\Big), \label{eq:MgMi} 
\end{equation}

\noindent where $m$ is the mass of a body, $E$ is the body's (negative) gravitational self-energy, $mc^2$ is its total mass-energy, and $\eta$ is a dimensionless constant for SEP violation \cite{Nordtvedt_1968b,Ken_LLR68}.
Any SEP violation is quantified by the parameter $\eta$: in fully-conservative, Lorentz-invariant theories of gravity \cite{Will_book93,Will-lrr-2006-3} the SEP parameter is related to the PPN parameters by $\eta = 4\beta - \gamma -3\equiv 4\bar\beta - \bar\gamma$. In general relativity $\gamma = \beta = 1$, so that $\eta = 0$ (see \cite{Williams-etal-2005,Will_book93,Will-lrr-2006-3}).

The quantity $E$ is the body's gravitational self-energy $(E< 0)$, which for a body $i$ is given by 
\begin{equation}
\left[{E  \over mc^2}\right]_i= - \frac{G}{2m_ic^2}\int_i d^3 x 
\rho_iU_i=- \frac{G}{2m_ic^2}\int_i d^3 x d^3 x' 
\frac{\rho_i({\bf r})\rho_i({\bf r}')}{| {\bf r} - {\bf r}'|}.        
\label{eq:omega}
\end{equation}
For a sphere with a radius $R$ and uniform density, $E /mc^2 = -3Gm/5Rc^2 = -0.3 v_E^2/c^2$, where $v_E$ is the escape velocity. Accurate evaluation for solar system bodies requires numerical integration of the expression of Eq.~(\ref{eq:omega}). Evaluating the standard solar model \cite{Ulrich_1982} results in $(E /mc^2)_\odot \sim -3.52 \times 10^{-6}$. Because gravitational self-energy is proportional to $m^2_i$ and also because of the extreme weakness of gravity, the typical values for the ratio $(E /mc^2)$ are $\sim 10^{-25}$ for bodies of laboratory sizes. Therefore, the experimental accuracy of 1 part in $10^{13}$ \cite{Adelberger_2001}, which is so useful for the WEP, is not sufficient to test how gravitational self-energy contributes to the inertial and gravitational masses of small bodies. To test the SEP one must consider planet-sized extended bodies, where the ratio in Eq.~(\ref{eq:omega}) is considerably higher.

\begin{wrapfigure}{R}{0.18\textwidth}
  \vspace{-15pt}
  \begin{center}
    \includegraphics[width=0.18\textwidth]{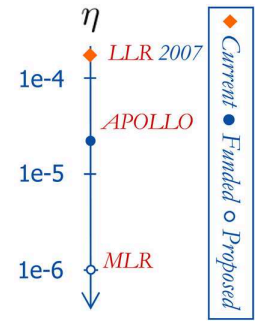}
  \end{center}
  \vspace{-00pt}
  \caption{Anticipated progress in the tests of the SEP \cite{Turyshev-etal-2007,Turyshev:2008dr}.}
  \vspace{-10pt}
\end{wrapfigure}

Currently, the Earth--Moon--Sun system provides the best solar system arena for testing the SEP. LLR experiments involve reflecting laser beams off retroreflector arrays placed on the Moon by the Apollo astronauts and by an unmanned Soviet lander \cite{Williams-etal-2004,Williams-etal-2005}. Recent solutions using LLR data give $(-0.8 \pm 1.3) \times 10^{-13}$ for any possible inequality in the ratios of the gravitational and inertial masses for the Earth and Moon.  This result, in combination with laboratory experiments on the WEP, yields a SEP test of $(-1.8 \pm 1.9) \times 10^{-13}$  that corresponds to the value of the SEP violation parameter of $\eta =(4.0 \pm 4.3) \times 10^{-4}$. 
In addition, using the recent Cassini result for the PPN parameter $\gamma$, PPN parameter $\beta$ is determined at the level of $\bar\beta=(1.2\pm1.1)\times 10^{-4}$ (see \cite{Williams-etal-2004} for details).

With the new Apache Point Observatory Lunar Laser-ranging Operations (APOLLO)\footnote{\label{ref:apollo}The Apache Point Observatory Lunar Laser-ranging Operations (APOLLO) is the new LLR station that was recently  built in New Mexico and  initiated operations in 2006.} facility \cite{Murphy-etal-2007,Williams-Turyshev-Murphy-2004}, LLR science has begun a renaissance.  APOLLO's 1-mm range precision will translate into order-of-mag\-nitude accuracy improvements in tests of the WEP and the SEP (leading to accuracy at levels of ${\Delta a}/{a}\lesssim 1\times10^{-14}$ and $\eta\lesssim 2\times10^{-5}$ respectively), in the search for variability of Newton's gravitational constant (see Sec.~\ref{sec:variation-G}), and in the test of the gravitational inverse-square law (see Sec.~\ref{sec:inv-sq-law}) on scales of the Earth-to-Moon distance (anticipated accuracy of $3\times10^{-11}$) \cite{Williams-Turyshev-Murphy-2004}.

The next step in this direction is interplanetary laser ranging \cite{laser-transponders-2006-1,laser-transponders-2006-2,Chandler-etal-2005,Turyshev-Williams-2007,Merkowitz-etal-2007} to, for example, a lander on Mars. Technology is available to conduct such measurements with a few-picosecond timing precision, which could translate into millimeter-class accuracies in ranging between the Earth and Mars. The resulting Mars laser ranging (MLR) experiment could (a) test the strong form of the EP with an accuracy at the $2\times10^{-6}$ level; (b) measure the PPN parameter $\gamma$ (see Sec.~\ref{sec:mod-grav}) with accuracy below the $10^{-6}$ level, and (c) test the gravitational inverse-square law at $\sim2$-AU distances with an accuracy of $1\times 10^{-14}$, thereby greatly improving the accuracy of the current tests \cite{Turyshev-Williams-2007}. MLR could also advance research in several areas of science including remote-sensing geodesic and geophysical studies of Mars.

Furthermore, with the recently demonstrated capabilities of reliable laser links over large distances (e.g., tens of millions of kilometers) in space \cite{laser-transponders-2006-1,laser-transponders-2006-2}, there is a strong possibility of improving the accuracy of gravity experiments with precision laser ranging over interplanetary scales \cite{Chandler-etal-2005,Turyshev-Williams-2007,Merkowitz-etal-2007}. The justification for such experiments is strong, the required technology is space-qualified, and some components have already flown in space. With MLR, the very best venue for gravitational physics will be expanded to interplanetary distances, representing an upgrade in both the scale and the precision of this promising technique.

The experiments described above are examples of the rich opportunities offered by the fundamental physics community to explore the validity of the EP. These experiments could potentially offer an improvement up to 5 orders of magnitude  over the  accuracy of the current tests of the EP.  Such experiments would dramatically enhance the range of validity for one of the most important physical principles or they could lead to a spectacular discovery.

\subsection{Tests of Local Lorentz Invariance: Search for Physics Beyond the Standard Model}
\label{sec:LLI}

Recently there has been an increase in activity in experimental tests of LLI, in particular light-speed isotropy tests. This increase is largely due to (a) advances in technology, which have allowed more precise measurements, and (b) the emergence of the Standard Model Extension (SME) as a framework for the analysis of experiments, which has provided new interpretations of LLI tests. None of the experiments performed to date has yet reported a violation of LLI, although the constraints on a putative violation have improved significantly.

LLI is an underlying principle of relativity, postulating that the outcome of a local experiment is independent of the velocity and orientation of the apparatus. 
To identify a violation it is necessary to have an alternative theory to interpret the experiment, and many have been developed. 
The Robertson-Mansouri-Sexl (RMS) framework \cite{Robertson-1949,Mansouri-Sexl-1977,Will-lrr-2006-3,Mattingly:2005re} is a well known kinematic test theory for parameterizing deviations from Lorentz invariance. In the RMS framework, there is assumed to be a preferred frame $\Sigma$ where the speed of light is isotropic. One usually analyzes the change in resonator frequency as a function of the Poynting vector direction with respect to the velocity of the lab in some preferred frame (as in \cite{Wolf-etal-2004,Tobar:2005vg}), typically chosen to be the cosmic microwave background.

The ordinary Lorentz transformations to other frames are generalized to
\begin{eqnarray}
t' = a^{-1}(t -\frac{\bf v \cdot {\bf x}}{c^2}), \qquad
{\bf x}' = d^{-1}{\bf x} - (d^{-1} - b^{-1})\frac{{\bf v}({\bf v}\cdot {\bf x})}{v^2} + a^{-1}{\bf v}t,
\label{eq:rms-framework}
\end{eqnarray}
where the coefficients $a, b, d$ are functions of the magnitude $v$ of the relative velocity between frames. This transformation is the most general one-to-one transformation that preserves rectilinear motion
in the absence of forces. In the case of special relativity, with Einstein clock synchronization, these coefficients reduce to $a = b^{-1} = \sqrt{1 - (v/c)^2}, d = 1$.  Many experiments, such as those that measure the isotropy of the one way speed of light \cite{Will-1992} or the propagation of light around closed loops, have observables that depend on $a, b, d$ but not on the synchronization procedure. Due to its simplicity, RMS has been widely used to interpret many experiments \cite{Mattingly:2005re}.
Most often, the RMS framework is used in situations where the velocity $v$ is small compared to $c$. We therefore expand $a, b, d$ in a power series in $v/c$:
\begin{eqnarray}
a = 1 + \alpha\frac{v^2}{c^2} + {\cal O}(c^{-4}) \qquad
b = 1 + \beta\frac{v^2}{c^2} + {\cal O}(c^{-4}) \qquad
d = 1 + \delta\frac{v^2}{c^2} + {\cal O}(c^{-4}). 
\end{eqnarray}
\noindent 
The RMS parameterizes a possible Lorentz violation by
a deviation of the parameters $(\alpha,\beta,\delta)$ from their special relativistic values $\alpha=-\frac{1}{2},~\beta=\frac{1}{2},~\delta=0$. 
These are typically grouped into three linear combinations that represent a measurement of (a) the isotropy of the speed of light or the orientation dependence ($P_{\rm MM} = \frac{1}{2}-\beta+\delta)$, measured in a Michelson-Morley (MM) experiment  \cite{Michelson-Morley-1887} and constrained by \cite{Muller-etal:2003,Wolf-etal-2004,Stanwix:2006jb} to $(9.4\pm 8.1) \times 10^{-11}$, (b) the boost dependence of the speed of light ($P_{\rm KT} = \beta- \alpha - 1)$, measured in a Kennedy-Thorndike (KT) experiment \cite{Kennedy-Thorndike-1932} and constrained by \cite{Wolf-etal-2003,Wolf-etal-2004} to $(3.1\pm 6.9) \times 10^{-7}$, and (c) the time dilation parameter ($P_{\rm IS} = |\alpha+\frac{1}{2}|$) measured in an Ives-Stillwell (IS) experiment \cite{Ives-Stilwell:1938} and constrained by \cite{Saathoff-etal:2003}
to $2.2\times10^{-7}$.  A test of Lorentz invariance was performed by comparing the resonance frequencies of two orthogonal cryogenic optical resonators subject to Earth's rotation over $\sim1$~yr. For a possible anisotropy of the speed of light $c$, the authors of Ref.~(\cite{Muller-etal:2003}) reported a constraint of $\Delta c/c=(2.6\pm1.7)\times10^{-15}$ which has since further improved \cite{Muller-etal:2007} by an additional order of magnitude. 

The RMS framework, however, is incomplete, as it says nothing about dynamics or about how given clocks and rods relate to fundamental particles. In particular, the coordinate transformation of Eq.~(\ref{eq:rms-framework}) only has meaning if we identify the coordinates with the measurements made by a particular set of clocks and rods. If we choose a different set of clocks and rods, the transformation laws may be different. Thus, it is not possible to compare the RMS parameters of two experiments that use physically different clocks and rods.
However, for experiments involving a single type of clock/rod and light, the RMS formalism is applicable and can be used to search for violations of Lorentz invariance in that experiment.\footnote{The RMS formalism can be made less ambiguous by placing it into a complete dynamical framework, such as the SME. It has been shown \cite{Kostelecky:2002hh} that the RMS framework can be incorporated into the SME.}

\begin{wrapfigure}{R}{0.46\textwidth}
  \vspace{-10pt}
  \begin{center}
    \includegraphics[width=0.46\textwidth]{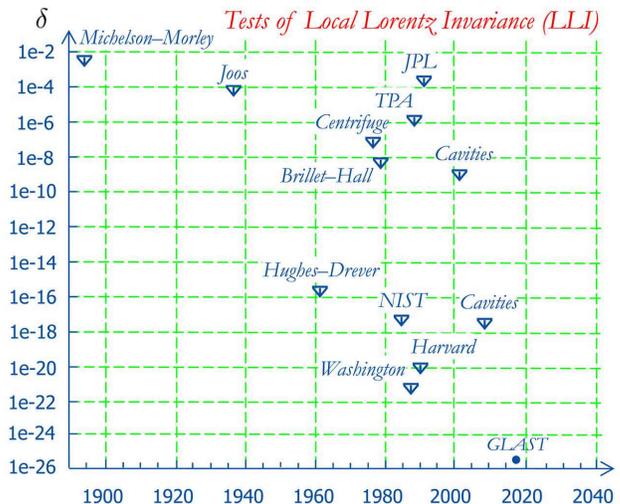}
  \end{center}
  \vspace{-10pt}
  \caption{The progress in the tests of the LLI since 1900s  \cite{Turyshev-etal-2007} and anticipated performance of the GLAST mission (developed jointly by the NASA and the DOE). 
}
\label{fig:LLI-tests}
  \vspace{-5pt}
\end{wrapfigure}

Limits on the violation of Lorentz symmetry are available from laser interferometric versions of the Michelson-Morley experiment, which compare the velocity of light, $c$ and the maximum attainable velocity of massive particles, $c_i$, up to $\delta \equiv
|c^2/c_{i}^2 - 1| < 10^{-9}$ \cite{Brillet-Hall-1979}. More accurate tests can be performed via the Hughes-Drever experiment \cite{Hughes-etal:1960,Drever-1961} by searching for a time dependence of the quadrupole splitting of nuclear Zeeman levels along Earth's orbit. This technique achieves an impressive limit of $\delta < 3 \times 10^{-22}$ \cite{Lamoreaux-etal:1986}. A recent reassessment of these results reveals that more stringent bounds can be reached, up to 8 orders of magnitude higher \cite{Kostelecky:1999mr}. The parameterized post-Newtonian parameter $\alpha_{3}$ can be used to set astrophysical limits on the violation of momentum conservation and the existence of a preferred reference frame. This parameter, which vanishes in general relativity can be accurately determined from the pulse period of pulsars and millisecond pulsars \cite{Will-lrr-2006-3}. The most recent results yields a limit on the PPN parameter $\alpha_3$  of $|\alpha_{3}| < 2.2 \times 10^{-20}$ \cite{Bell:1995ax}. 

Since the discovery of the cosmological origin of gamma-ray bursts (GRBs), there has been growing interest in using these transient events to probe the quantum gravity energy scale ranging from $10^{16}$ to $10^{19}$~GeV, up to the Planck mass scale. This energy scale can manifest itself through a measurable modification in the electromagnetic radiation dispersion relation for high-energy photons originating from cosmological distances. 
The Gamma-ray Large Area Space Telescope (GLAST)\footnote{The Gamma-ray Large Area Space Telescope (GLAST), please see website: {\tt http://glast.gsfc.nasa.gov/}.} \cite{Ritz-etal:2007} is expected to improve the LLI tests by several orders of magnitude potentially reaching an accuracy at the level of $\delta \simeq 10^{-26}$ (see Fig.~\ref{fig:LLI-tests}) \cite{Mattingly:2005re,Ritz-etal:2007}. GLAST will measure the cosmic gamma-ray flux ranging from 20 MeV to $>$ 300 GeV, with supporting measurements for gamma-ray bursts from 8 keV to 30 MeV.  Launched in 2008, GLAST has opened a new and important window on a wide variety of phenomena, including black holes and active galactic nuclei, the optical-ultraviolet extragalactic background light, gamma-ray bursts, the origin of cosmic rays and supernova remnants, and searches for hypothetical new phenomena such as supersymmetric dark matter annihilations and Lorentz-invariance violation. 

The Standard Model coupled to general relativity is thought to be the effective low-energy limit of an underlying fundamental theory that unifies gravity and particle physics at the Planck scale. This underlying theory may well include Lorentz violation \cite{Colladay:1996iz,Colladay:1998fq,Coleman:1997xq,Coleman:1998ti} which could be detectable in space-based experiments \cite{Kostelecky:1994rn}.  Lorentz symmetry breaking due to non-trivial solutions of string field theory was first discussed in Ref. \cite{Kostelecky-Samuel:1989a,Kostelecky-Samuel:1989b}. These solutions arise from the string field theory of open strings and may have implications for low-energy physics. For instance, assuming that
the contribution of Lorentz-violating interactions to the vacuum energy is about half of the critical density implies that feeble tensor-mediated interactions in the range of $ \sim 10^{-4}$~m should exist \cite{Bertolami:1997,Bertolami-Paramos-Turyshev-2007}. Also, violations of Lorentz invariance may imply in a breaking of the fundamental charge-parity-time CPT symmetry of local quantum field theories \cite{Kostelecky-Samuel:1991,Kostelecky:1995qk,Kostelecky:2000hz,Kostelecky:1999mu}. Quite remarkably, this can be experimentally verified in neutral-meson \cite{Colladay:1994cj,Colladay:1995qb} experiments, Penning-trap measurements \cite{Bluhm:1997ci,Bluhm:1997qb} and hydrogen-antihydrogen spectroscopy \cite{Bluhm:1998rk,Russell:2008uz}. This spontaneous breaking of CPT symmetry allows for an explanation of the baryon asymmetry of the Universe: In the early Universe, after the breaking of the Lorentz and CPT symmetries, tensor-fermion interactions in the low-energy limit of string-field theories gave rise to a chemical potential that created in equilibrium a baryon-antibaryon asymmetry in the presence of baryon number violating interactions \cite{Bertolami:1996cq,Russell:2008uz}. The development of SME has inspired a new wave of experiments designed to explore the uncharted regions of the Lorentz-violating parameter space.

\begin{wrapfigure}{R}{0.46\textwidth}
  \vspace{-20pt}
  \begin{center}
    \includegraphics[width=0.46\textwidth]{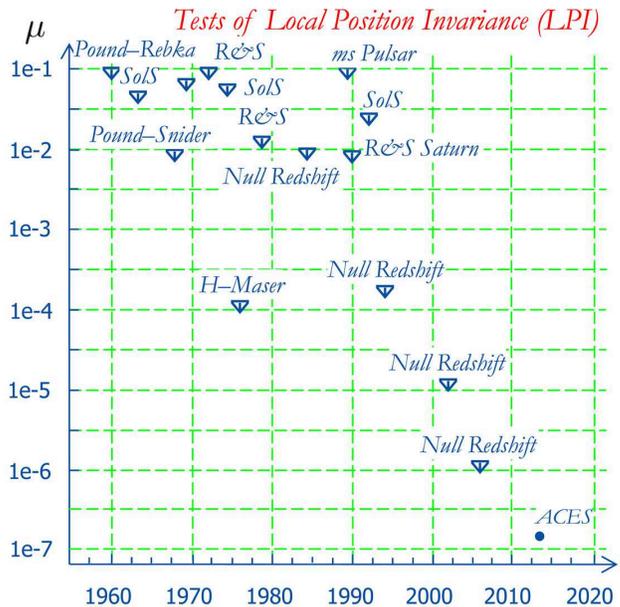}
  \end{center}
  \vspace{-10pt}
  \caption{The progress in the tests of the LPI since the early 1950s  \cite{Turyshev-etal-2007} and anticipated performance of the ESA's ACES mission. Here `SolS', for the tests with solar spectra, `R\&S' is that using rockets and spacecraft, and `Null Redshift' is for comparison of different atomic clocks (see \cite{Will-lrr-2006-3}).
}
  \vspace{-10pt}
\label{fig:LPI-tests}
\end{wrapfigure}

If one adds to the Standard Model the appropriate terms involving operators for Lorentz-invariance violation \cite{Stecker:2001vb}, the result is the SME, which has provided a phenomenological framework for testing Lorentz invariance
\cite{Kostelecky:1995qk,Kostelecky:2000hz,Kostelecky:1999mu,Kostelecky-Samuel:1989a,Kostelecky-Samuel:1989b}, and which has also suggested a number of new tests of relativistic gravity in the solar system \cite{Bailey:2006fd}. Compared with their ground-based analogs, space-based experiments in this area can provide improvements by as much as six orders of magnitude.  Several general reviews of the SME and corresponding efforts are available (for review, see \cite{Mattingly:2005re,Bluhm:2005uj,Kostelecky-CPT:1999,Kostelecky-CPT:2002,Kostelecky-CPT:2005}).   Recent studies of the ``aether theories'' \cite{Jacobson:2000xp,Foster:2005dk,Jacobson:2004ts} have shown that these models are naturally compatible with general relativity \cite{Will-lrr-2006-3}, but that they predict several non-vanishing Lorentz-violation parameters that could be measured in experiment. The authors of Ref.~\cite{Kostelecky:2008ts} tabulate experimental results for coefficients for Lorentz and CPT violation in the minimal SME formalism and report the attained sensitivities in the matter and photon sectors.

Searches for extensions of special relativity on a space-based platforms are known as ``clock-comparison'' tests. Such tests involve operating two or more high-precision clocks simultaneously and comparing their rates correlated with orbit parameters, such as velocity relative to the CMB and to position in a gravitational environment.  The SME allows for the possibility that comparisons of the signals from different clocks will yield very small differences that can be detected in experiment. For present-day results, see Ref.~\cite{Kostelecky:2008ts} which provides a summary of experimental results for the coefficients for Lorentz and CPT violation within the minimal SME formalism.

Tests of special relativity and the SME were proposed by the Superconducting Microwave Oscillator (SUMO) group, those from the Primary Atomic Reference Clock in Space (PARCS) \cite{Heavner-etal:2001,Ashby-2002,Sullivan-etal:2005} and the Rubidium Atomic Clock Experiment (RACE) \cite{Gibble-etal:2004} originally slated for operation on the International Space Station (ISS) in the 2005-07 time frame. SUMO, a cryogenic cavity experiment \cite{Lipa-etal:2005}, was to be linked with PARCS to provide differential red-shift and Kennedy-Thorndike measurements and improved local oscillator capability \cite{Sullivan-etal:2005}. Unfortunately, for programmatic reasons the development of these experiments was canceled by NASA in 2004. Presently, an experiment, known as the Atomic Clock Ensemble in Space (ACES), is aiming to perform important tests of SME.  ACES is a European mission \cite{Cacciapuoti-etal:2007} in fundamental physics that will operate atomic clocks in the microgravity environment of the International Space Station (ISS) with fractional frequency stability and an accuracy of a few parts in 10$^{16}$.  ACES is jointly funded by the ESA and the CNES and is being prepared for a 2013-14 flight to the ISS \cite{Salomon-etal:2007} for a planned mission duration of 18 months \cite{Turyshev-etal-2007}.

Optical clocks offer an improved possibility of testing the time variations of fundamental constants at a high level of accuracy  \cite{Marion:2002iw,Bize-etal:2003,Fischer-etal:2004,Peik-etal:2004,Fortier-etal-2007} (also see \cite{Turyshev-etal-2007} and references therein).  Interestingly, such measurements complement tests of the LLI \cite{Wolf-Petit:1997}  and of the UFF to experimentally establish the validity of the EP. The universality of the gravitational red-shift can be tested at the same accuracy level by two optical clocks in free flight in a varying gravitational potential. The constancy and isotropy of the speed of light can be tested by continuously comparing a space clock with a ground clock.  Optical clocks orbiting the Earth, combined with a sufficiently accurate time and frequency transfer link, can improve present results by more than three orders of magnitude.

There is a profound connection between cosmology and possible Lorentz symmetry violation \cite{Kostelecky-2004,Dvali-Pujolas-Redi-2007}. Spontaneous breaking of the Lorentz symmetry implies that there exists an order parameter with a non-zero expectation value that is responsible for the effect. For spontaneous Lorentz symmetry breaking one usually assumes that sources other than the familiar matter density are responsible for such a violation.  However, if Lorentz symmetry is broken by an extra source, the latter must also affect the cosmological background.  Therefore, in order to identify the mechanism of such a violation, one must look for traces of similar symmetry breaking in cosmology, for instance, in the CMB data.\footnote{Analyses of the CMB for Lorentz violation have already begun \cite{Feng:2006dp,Kostelecky-Mewes-2007}. This provides a systematic classification of all operators for Lorentz violation and uses polarimetric observations of the cosmic microwave background to search for associated effects. Lorentz symmetry violation can also have important implication for cosmology via CPT violation and baryogenesis \cite{Bertolami:1996cq,Carroll:2005dj,Li:2006ss}.} In other words, were a violation of the Lorentz symmetry discovered in experiments but not supported by observational cosmology data, such a discrepancy would indicate the existence of a novel source of symmetry breaking.  This source would affect the dispersion relation of particles and the performance of the local clocks, but it would leave no imprint on the cosmological metric.  Such a possibility emphasizes the importance of a comprehensive program to investigate all possible mechanisms of breaking of the Lorentz symmetry, including those accessible by experiments conducted in space-based laboratories.

\subsection{Tests of the Local Position Invariance}
\label{sec:LPI}

Einstein predicted the gravitational redshift of light from the EP in 1907, but the redshift is very difficult to measure astrophysically. Given that both the WEP and the LLI postulates have been tested with great accuracy, experiments concerning the universality of the gravitational red-shift measure the level to which the LPI holds. Therefore, violations of the LPI would imply that the rate of a free falling clock would be different when compared with a standard one, for instance on the Earth's surface. The accuracy to which the LPI holds as an invariance of Nature can be parameterized through $ \Delta \nu / \nu = (1 + \mu) U / c^2$. 

The first observation of the gravitational redshift was the measurement of the shift in the spectral lines from the white dwarf star Sirius B by Adams in 1925. Although this measurement, as well as later measurements of the spectral shifts on other white dwarf stars, agreed with the prediction of relativity, the shift might stem from some other cause; hence, experimental verification using a known terrestrial source is preferable. The effect was conclusively tested by Pound \& Rebka's 1959 experiment. 

The Pound--Rebka experiment was one of the first precision experiments testing general relativity; it further verified the effects of gravity on light by testing the universality of the gravity-induced frequency shift, $\Delta \nu$, that follows from the WEP:
${\Delta \nu / \nu} = {g h / c^2} = (2.57 \pm 0.26) \times 10^{-15},$ where $g$ is the acceleration of gravity and $h$ is the height of fall \cite{Pound-Rebka-1959,Pound-Rebka-1960}. The test of LPI resulted in a limit of $\mu \simeq 10^{-2}$ \cite{Pound-Snider-1964}. The experiment was based on M\"ossbauer-effect measurements between sources and detectors spanning the 22.5 m tower in the Jefferson Physical Laboratory at Harvard University.  

In 1976, an accurate verification of the LPI was performed by Vessot and collaborators who compared the frequencies of two Hydrogen masers, one being on Earth and another one on a sub-orbital rocket.  The resulting Gravity Probe A experiment \cite{Vessot-etal-1980} exploited the much higher ``tower'' enabled by space. A sub-orbital Scout rocket carried a Hydrogen maser to an altitude of 10273 km, and a novel telemetry scheme allowed comparison with Hydrogen masers on the ground.  The Vessot and Levine \cite{Vessot-Levine-1979,Vessot-etal-1980} experiment verified that the fractional change in the measured frequencies is consistent with general relativity to the $10^{-4}$ level, confirming Einstein's prediction to 70 ppm and thereby establishing a limit of $ \vert \mu \vert < 2 \times 10^{-4}$. More than 30 years later, this remained the most precise measurement of the gravitational red-shift \cite{Will-lrr-2006-3}. The universality of this redshift has also been verified by measurements involving other types of clocks. Currently the the most stringent bound on possible violation of the LPI is $ \vert \mu \vert < 2.1 \times 10^{-5}$ \cite{Bauch-Weyers-2002}. The accuracy of few parts in 10$^6$ in differential measurements of $\mu$ was reported in \cite{Ashby-etal-2007}. The ESA's ACES mission is expected to improve the results of the LPI tests (see Fig.~\ref{fig:LPI-tests}). With the full accuracy of ground and space clocks at the $10^{-16}$ level or better, Einstein's effect can be tested with a relative uncertainty
of $\mu \simeq 2 \times 10^{-6}$, yielding improvement of up to a factor 35 with respect to the previous experiment \cite{Salomon-etal:2007}.

As mentioned above, gravitational redshift has been measured both in the laboratory \cite{Pound-Rebka-1959,Pound-Rebka-1960,Pound-Snider-1964} and by using astronomical observations \cite{Greenstein-Oke-Shipman-1971,Barstow:2005mx}. Gravitational time dilation in the Earth's gravitational field has been measured numerous times using atomic clocks \cite{Vessot-etal-1980,Bauch-Weyers-2002}, and ongoing validation is provided as a side-effect of the operation of the Global Positioning System (GPS) \cite{Ashby-lrr-2003-1}. Tests in stronger gravitational fields are provided by the observation of binary pulsars \cite{Stairs-lrr-2003-5,Kramer:2004gi,Kramer-etal-2006}. All results are in agreement with general relativity \cite{Will-lrr-2006-3}; however, at the current level of accuracy, these observations cannot distinguish between general relativity and other metric theories that preserve the EP.

\subsection{Search for Variability of the Fundamental Constants}
\label{sec:fest-fund-const}

Dirac's 70 year old idea of cosmic variation of physical constants has been revisited with the advent of models unifying the forces of nature based on the symmetry properties of possible extra dimensions, such as the Kaluza-Klein-inspired theories, Brans-Dicke theory, and supersymmetry models.  Alternative theories of gravity \cite{Will-lrr-2006-3} and theories of modified gravity \cite{Bertolami-Paramos-Turyshev-2007} include cosmologically evolving scalar fields that lead to variability of the fundamental constants. It has been  hypothesized that a variation of the cosmological-scale factor with epoch could lead to temporal or spatial variation of the physical constants, specifically the gravitational constant, $G$, the fine-structure constant, $\alpha = e^2/\hbar c\simeq1/137.037$, and the electron-proton mass ratio ($m_{\rm e}/m_{\rm p}$) \cite{Wilczek:2007}. 

In general, constraints on the variation of fundamental constants can be derived from a number of gravitational measurements, such as the test of the UFF, the motion of the planets in the solar system, and stellar and galactic evolutions. The constraints are based on the comparison of two time scales, the first (gravitational time) dictated by gravity (e.g., ephemeris, stellar ages, etc.) and the second (atomic time) determined by a non-gravitational system (e.g. atomic clocks, etc.) \cite{Canuto-Goldman-1982-1,Canuto-Goldman-1982-2}. For instance, planetary and spacecraft ranging, neutron star binary observations, and paleontological and primordial nucleosynthesis data allow one to constrain the relative variation of $G$ \cite{Uzan-2003}. Many of the corresponding experiments could reach a much higher precision if performed in space. 

\subsubsection{Fine-Structure Constant} 
\label{sec:variation-alpha}

The current limits on the evolution of $\alpha$ are established by laboratory measurements, studies of the abundances of radioactive isotopes and those of fluctuations in the CMB and other cosmological constraints (for a review see \cite{Uzan-2003}).  There exist several types of tests that are based, for instance, on geological data (e.g., measurements made on the nuclear decay products of old meteorites) and on measurements (of astronomical origin) of the fine structure of absorption and emission spectra of distant atoms (e.g., the absorption lines of atoms on the line-of-sight of quasars at high redshift -- an important observational technique which is outside the scope of this review). 
Laboratory experiments are based on the comparison either of different atomic clocks or of atomic clocks with ultra-stable oscillators. They also have the advantage of being more reliable and reproducible, thus allowing better control of the systematics and better statistics compared with other methods. Their evident drawback is their short time-scales, which fixed by the fractional stability of the least-precise standards. These time-scales are usually of order of a month to a year, so the obtained constraints are restricted to the instantaneous variation observed today. However, the shortness of the time scales is compensated by a much higher experimental sensitivity. All of these kinds of tests all depend on the value of $\alpha$.

The best measurement of the constancy of $\alpha$ to date is that provided by the Oklo phenomenon; it sets the following (conservative) limits on the variation of $\alpha$ over a period of two billion years \cite{Shlyakhter-1976,Shlyakhter-1982,Damour-Dyson-1996,Fujii-etal-2000,Fujii-etal-2002}: $-0.9\times10^{-7}<\alpha^{\rm Oklo}/\alpha^{\rm today}-1<1.2\times10^{-7}$.
Converting this result into an average time variation, one finds
\begin{equation}
-6.7\times10^{-17}~{\rm yr}^{-1}<\frac{\dot\alpha}{\alpha}<5\times10^{-17}~{\rm yr}^{-1}.
\end{equation}

\noindent Note that this variation is a factor of $\sim 10^7$ smaller than the Hubble scale, which is itself $\sim10^{-10}$~yr$^{-1}$. Comparably stringent limits were obtained using the
Rhenium 187 to Osmium 187 ratio in meteorites \cite{Olive-etal-2004}, which yielded an upper bound of $\dot\alpha/\alpha=(8\pm8)\times10^{-7}$
over $4.6\times10^9$ years. Laboratory limits were also obtained
from the comparison, over time, of stable atomic clocks. More precisely, given that $v/c\sim\alpha$ for electrons in the first Bohr orbit, direct measurements of the variation of $\alpha$ over time can be made by comparing the frequencies of atomic clocks that rely on different atomic transitions. The upper bound on the variation of $\alpha$ using such methods is $\dot\alpha/\alpha= (-0.9 \pm 2.9) \times10^{-15}$~yr$^{-1}$ \cite{Bize-etal:2003,Marion:2002iw,Fischer-etal:2004}. With the full accuracy of ground and space clocks at the $10^{-16}$ level or better, the ESA's ACES mission will be able to measure time variations of the fine structure constant at the level of $\simeq10^{-16}$~yr$^{-1}$ \cite{Salomon-etal:2007}.

There is a connection between the variation of the fundamental constants and a violation of the EP; in fact, the former almost always implies the latter. For example, should there be an ultra-light scalar particle, its existence would lead to variability of fundamental constants, such as $\alpha$ and $m_{\rm e}/m_{\rm p}$. Because  masses of nucleons are $\alpha$-dependent, by coupling to nucleons this particle would mediate an isotope-dependent long-range force \cite{Damour_Polyakov_94b,Damour-2002,Dvali-Zaldarriaga-2002,Uzan-2003,Dent-2007}. The strength of the coupling is within a few of orders of magnitude of existing experimental bounds for such forces; thus, the new force could potentially be measured in precision tests of the EP. Therefore, the existence of a new interaction mediated by a massless (or very low-mass) time-varying scalar field would lead both to the variation of the fundamental constants and to violation of the WEP, ultimately resulting in observable deviations from general relativity.

Following the arguments above, for macroscopic bodies, one would expect that their masses depend on all the coupling constants of the four known fundamental interactions, which have profound consequences concerning the motion of a body. In particular, because the $\alpha$-dependence is {\it a priori} composition-dependent, any variation of the fundamental constants entails a violation of the UFF \cite{Uzan-2003}. This allows comparison of the ability of two classes of experiments -- namely clock-based and EP-testing experiments -- to search for variation of the parameter $\alpha$ in a model-indepen\-dent way \cite{Nordtvedt-2002}. EP experiments have been superior performers. Thus, analysis of the frequency ratio of the 282-nm $^{199}$Hg$^+$ optical clock transition to the ground-state hyperfine splitting in $^{133}$Cs was recently used to place a limit on its fractional variation of $\dot\alpha/\alpha \leq 1.3\times 10^{-16}~{\rm yr}^{-1}$ \cite{Fortier-etal-2007}. At the same time, the current accuracy of the EP tests \cite{Williams-etal-2005} already constrains the variation as $\Delta \alpha/\alpha \leq 10^{-10}\Delta U/c^2$, where $\Delta U$ is the change in the gravity potential. Therefore, for ground-based experiments (for which the variability in the gravitational potential is due to the orbital motion of the Earth), the quantity $U_{\tt sun}/c^2$ varies by $1.66\times10^{-10}$ over one year, so a ground-based clock experiment must therefore be able to measure fractional frequency shifts between clocks to a precision of 1 part in $10^{20}$ in order to compete with EP experiments on the ground \cite{Nordtvedt-2002}. 

However, sending atomic clocks on a spacecraft to within a few solar radii of the Sun where the gravitational potential grows to $10^{-6} c^2$ could be a competitive experiment if the relative frequencies of different on-board clocks could be measured to a precision better than 1 part in $10^{16}$. Such an experiment would allow for a direct measurement of any $\alpha$-variation, thus further motivating the development of  space-qualified clocks. With their accuracy poised to surpass the $10^{-17}$ level in the near future, optical clocks may be able to provide the needed capabilities  to directly test the variability of the fine-structure constant \cite{Turyshev-etal-2007}. 

{\it SpaceTime} is a proposed atomic-clock experiment designed to search for a variation of the fine-structure constant with a detection sensitivity of $\dot\alpha/\alpha\sim 10^{-20}~{\rm yr}^{-1}$ and will be carried out on a spacecraft that flies to within six solar radii of the sun \cite{Maleki-Prestage-2007}. The test relies on an instrument utilizing a tri-clock assembly that consists of three trapped-ion clocks based on mercury, cadmium, and ytterbium ions that are placed in the same vacuum, thermal and magnetic field environment.  Such a configuration allows for a differential measurement of the frequency of the clocks, and the cancellation of perturbations common to the three. For alkali atoms, the sensitivity of different clocks, based on atoms of different $Z$, to a change in the fine structure constant display specific signatures.  In particular, the Casimir correction factor, $F(\alpha Z)$, leads to the differential sensitivity in the alkali microwave hyperfine clock transition frequencies. As a result, different atomic systems with different $Z$ display different frequency dependencies on a variation of $\alpha$ through the $\alpha Z$ dependent terms.  A direct test for a time variation of $\alpha$ can then be devised through a comparison of two clocks, based on two atomic species with different atomic number, $Z$.  This is a key feature of the SpaceTime instrument that in conjunction with the individual sensitivity of each atomic species to an $\alpha$-variation, can produce clear and unambiguous results. Observation of any frequency drift between the three pairs of the clocks in response to the change in gravitational potential, as the tri-clock instrument approaches the sun, would signal a variation in $\alpha$.

\begin{wrapfigure}{R}{0.16\textwidth}
  \vspace{-10pt}
  \begin{center}
    \includegraphics[width=0.16\textwidth]{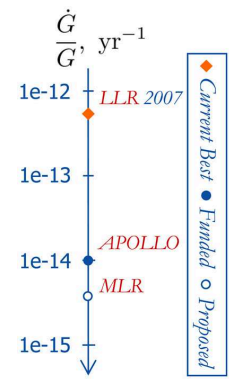}
  \end{center}
  \vspace{-10pt}
  \caption{Anticipated progress in the tests of possible variability in the gravitational constant \cite{Turyshev-etal-2007,Turyshev:2008dr}.}
  \vspace{-10pt}
\label{fig:future-G-dot-tests}
\end{wrapfigure}

Clearly a solar fly-by on a highly-eccentric trajectory with very accurate clocks and inertial sensors makes for a compelling relativity test.  A potential use of highly-accurate optical clocks in such an experiment would likely lead to additional accuracy improvement in the tests of $\alpha$ and $m_{\rm e}/m_{\rm p}$, thereby providing a good justification for space deployment \cite{Schiller-etal-2006,Turyshev-etal-2007}. The resulting space-based laboratory experiment could lead to an important discovery.

\subsubsection{Gravitational Constant} 
\label{sec:variation-G}

A possible variation of Newton's gravitational constant $G$ could be related to the expansion of the universe depending on the cosmological model considered. Variability in $G$ can be tested in space with a much greater precision than on Earth \cite{Williams-etal-2004,Uzan-2003}.  For example, a decreasing gravitational constant, $G$, coupled with angular momentum conservation is expected to increase a planet's semimajor axis, $a$, as $\dot a/a=-\dot G/G$.  The corresponding change in orbital phase grows quadratically with time, providing for strong sensitivity to the effect of $\dot G$.

Currently, space-based experiments using lunar and planetary ranging measurements are the best means of searching for very small spatial or temporal gradients in the values of $G$ \cite{Williams-etal-2004,Williams-etal-2005}.  Thus, recent analysis of LLR data strongly limits such variations and constrains a local scale ($\sim$1~AU) expansion of the solar system as $\dot a/a=-\dot G/G =-(5\pm6) \times 10^{-13}$ yr$^{-1}$, including the expansion resulting from cosmological effects \cite{Williams-etal-2007}. Interestingly, the achieved accuracy in $\dot G/G$ implies that, if this rate is representative of our cosmic history, then $G$ has changed by less than 1\% over the 13.4 Gyr age of the universe.

The ever-extending LLR data set and the increase in the accuracy of lunar ranging (i.e., APOLLO) could lead to significant improvements in the search for variability of Newton's gravitational constant; an accuracy at the level of $\dot G/G\sim 1\times10^{-14}~{\rm yr}^{-1}$ is feasible with LLR \cite{Turyshev-Williams-2007}. High-accuracy timing measurements of binary and double pulsars could also provide a good test of the variability of the gravitational constant \cite{Nordtvedt-2002,Kramer-etal-2006}. A preliminary analysis of the the accuracy achievable  with MLR indicates that $\dot G/G$ could be determined with an accuracy at the level $\dot G/G\sim 3\times10^{-15}~{\rm yr}^{-1}$ (limited by the asteroids and the lifetime of the experiment). Furthermore, MLR could also determine the solar mass loss with accuracy at the level of $\dot M_\odot/M_\odot\sim 3\times10^{-14}~{\rm yr}^{-1}$ (theoretically expected value is $7\times10^{-14}~{\rm yr}^{-1}$). Fig.~\ref{fig:future-G-dot-tests} shows the anticipated progress in the tests of possible variability in the gravitational constant. 

\subsection{\label{sec:inv-sq-law}Search for New Physics via Tests of the Gravitational Inverse-Square Law}  

Many modern theories of gravity, including the string, supersymmetry, and brane-world theories, suggest that new physical interactions will appear at short ranges.  This may happen because at sub-millimeter distances new dimensions can exist, thereby changing the gravitational inverse-square law \cite{ADD-1998,ADD-1999} (for a review of experiments, see \cite{Adelberger-Heckel-Nelson-2003}). Similar forces that act at short distances are predicted in supersymmetric theories with weak scale compactifications \cite{AnDD-1998}, in some theories with very low energy supersymmetry breaking \cite{Dimopoulos-Giudice-1998}, and also in theories of very low quantum gravity scale \cite{Sundrum-1999,Dvali-etal-1998}. These multiple predictions provide strong motivation for experiments that would test for possible deviations from Newton's gravitational inverse-square law at very short distances, notably on millimeter-to-micrometer ranges.

An experimental confirmation of new fundamental for\-ces would provide an important insight into the physics beyond the Standard Model. A great deal of interest in the subject was sparked after the 1986 claim of evidence for an intermediate-range interaction with sub-gravitational strength \cite{Fischbach-Talmadge-1998}, leading to a wave of new experiments.

In its simplest versions, a new interaction (or a fifth force) would arise from the exchange of a light boson coupled to matter with a strength comparable to gravity. Planck-scale physics could give origin to such an interaction in a variety of ways, thus yielding a Yukawa-type modification in the interaction energy between point-like masses. This new interaction can be derived, for instance, from extended supergravity theories after dimensional reduction \cite{Scherk}, compactification of $5$-dimensional generalized Kaluza-Klein theories including gauge interactions at higher dimensions \cite{Bars-Visser-1986}, and also from string/M-theory. In general, the interaction energy, $V(r)$, between two point masses $m_1$ and $m_2$ can be expressed in terms
of the gravitational interaction as
{}
\begin{equation} 
V(r) = -  {G_{\infty}m_{1}m_{2} \over r}\big(1 +
\alpha\,e^{-r/\lambda}\big), \label{eq:2.1} 
\end{equation}
\noindent where $r = \vert {\bf r}_2 - {\bf r}_1 \vert$ is the distance between the masses, $G_{\infty}$ is the gravitational coupling for $r \rightarrow \infty$ and $\alpha$ and $\lambda$ are respectively the strength and range of the new interaction. Naturally, $G_{\infty}$ has to be identified with Newton's gravitational constant and the gravitational coupling becomes dependent on $r$. Indeed, the force associated with Eq.~(\ref{eq:2.1}) is given by: ${\bf F}(r)={\vec \nabla} V(r)=  
 - G(r)m_{1}m_{2}\,\hat {\bf r}/r^2$,  where  
$G(r) = G_{\infty}\big[1 + \alpha\,(1 + r/\lambda)e^{-r/\lambda}\big]$.

A new interaction results from an assumption that the coupling $\alpha$ is not an universal constant, but instead a composition-dependent parameter 
\cite{Lee-Yang-1955}. This comes about if one considers that the new bosonic field couples to the baryon number $B = Z + N$, which is the sum of protons and neutrons. Hence the new interaction between masses with baryon numbers $B_1$ and $B_2$ can be expressed through a new fundamental constant, $f$, as 
$V(r) = - f^{2}({B_{1}B_{2} / r})e^{-r/\lambda},$ such that the constant $\alpha$ can be written as
$\alpha= - \sigma [{B_{1} / \mu_{1}}][{B_{2} / \mu_{2}}], $
with $\sigma = f^{2}/G_{\infty}m_{\rm H}^{2}$ and $\mu_{1,2} = m_{1,2}/m_{\rm H}$, $m_{\rm H}$ being the hydrogen mass. Therefore, in a Galileo-type experiment a difference in acceleration exists between the masses $m_1$ and $m_2$, given by
\begin{equation} 
{\bf a}_{12}= \sigma\left[{B \over \mu}\right]_\oplus
\left(\left[{B_{1} \over \mu_{1}}\right]- \left[{B_{2} \over
\mu_{2}}\right]\right){\bf g}, \label{eq:2.6} 
\end{equation}
\noindent where {$\bf g$} is the field strength of the Earth's
gravitational field.

Several experiments (see, for instance, Refs. \cite{Fischbach-Talmadge-1998} for a list of the most relevant) studied the parameters of a new interaction based on the idea of a composition-dependence differential acceleration, as described in Eq.~(\ref{eq:2.6}), and other composition-independent effect. The current experimental status is essentially compatible with the predictions of Newtonian gravity, in both composition-independent or -dependent setups. The bounds on parameters $\alpha$ and $\lambda$ are summarized below 
\begin{itemize}
\item[--] Laboratory experiments devised to measure deviations from the inverse-square law are most sensitive in the range $10^{-2}~{\rm m} \lesssim \lambda \lesssim 1~{\rm m}$, constraining $\alpha$ to be smaller than about $10^{-4}$;

\item[--] Gravimetric experiments sensitive in the range of $10~{\rm m} \lesssim \lambda \lesssim 10^{3}~{\rm m}$ indicate that $\alpha \lesssim 10^{-3}$;

\item[--] Satellite tests probe the ranges of about  $10^{5}~{\rm m} \lesssim \lambda \lesssim 10^{7}~{\rm m}$ suggest that $\alpha \lesssim 10^{-7}$;

\item[--] Analysis of the effects of the inclusion of scalar fields into the stellar structure yields a bound in the range $10^8~{\rm m} \lesssim \lambda \lesssim 10^{10}~{\rm m}$, limiting $\alpha$ to be smaller than approximately $10^{-2}$.
\end{itemize}

Recent ground-based torsion-balance experiments \cite{Kapner:2006si} tested the gravitational inverse-square law at separations between 9.53 mm and $55~\mu$m, probing distances less than the dark-energy length scale $\lambda_d =\sqrt[4]{\hbar c/ u_d}\approx 85~\mu$m, and with an energy density $u_d\approx3.8~{\rm keV/cm}^3$.  It was found that the inverse-square law holds down to a length scale of $56~\mu$m and that an extra dimension must measure less than $44~\mu$m (similar results were obtained by \cite{Tu-etal-2007}).  These results are important, in that they signify the fact that modern experiments reached the level at which dark-energy physics can be tested in a laboratory setting; they also provide a new set of constraints on new forces \cite{Adelberger-etal-2007}, making these experiments relevant and competitive with particle physics research. Also, recent laboratory experiments testing Newton's second law for small accelerations \cite{Schlamminger-etal-2007,Gundlach-etal-2007} provided useful constraints relevant to understanding several current astrophysical puzzles.  New experiments are being designed to explore length scales below 5 $\mu$m \cite{Weld-Xia-Cabrera-Kapitulnik-2008}. 

Sensitive experiments searching for weak forces invariably require soft suspension for the measurement degree of freedom.  A promising soft suspension with low dissipation is superconducting magnetic levitation.  Levitation in 1-{\sl g}, however, requires a large magnetic field, which tends to couple to the measurement degree of freedom through metrology errors and coil nonlinearity, as well as to stiffen the mode.  The high magnetic field will also make suspension more dissipative.  The situation improves dramatically in space.  The {\sl g}-level is reduced by five to six orders of magnitude, so the test masses can be supported with weaker magnetic springs, which permits the realization of both the lowest resonance frequency and the lowest dissipation.  The microgravity conditions also allow for an improved design of the null experiment, free from the geometric constraints of the torsion balance.

The Inverse-Square Law Experiment in Space (ISLES) is a proposed experiment whose objective is to perform a highly accurate test of Newton's gravitational law in space \cite{Paik-etal-2004-1,Paik-etal-2004-2,Paik-etal-2007}. ISLES combines the advantages of the microgravity environment  with superconducting accelerometer technology to improve the current ground-based limits in the strength of violation \cite{Chiaverini-etal-2003} by four to six orders of magnitude in the range below $100~\mu$m.  The experiment will be sensitive enough to probe large extra dimensions down to $5~\mu$m and also to probe the existence of the axion\footnote{The axion is a hypothetical elementary particle postulated by Peccei-Quinn theory in 1977 to resolve the strong-CP problem in quantum chromodynamics (QCD); see details in Ref.~\cite{Peccei-Quinn:1977a,Peccei-Quinn:1977b,Weinberg:1978,Wilczek:1978}.} which---if it exists---is expected to violate the inverse-square law in the range accessible by ISLES.

The recent theoretical ideas concerning new particles and new dimensions have reshaped the way we think about the universe. Thus, should the next generation of experiments detect a force violating the inverse-square law, such a discovery would imply the existence of (a) an extra spatial dimension, (b) a massive graviton, or (c) the presence of a new fundamental interaction \cite{Adelberger-Heckel-Nelson-2003,Adelberger-etal-2007}.   

Although investigators have devoted much attention to the behavior of gravity at short distances, it is possible that tiny deviations from the inverse-square law occur at much larger distances. In fact, there is a possibility that non-compact extra dimensions could produce such deviations at astronomical distances \cite{Dvali-Gruzinov-Zaldarriaga-2003} (for discussion see Sec.~\ref{sec:mod-grav}).

By far the most stringent constraints on a test of the inverse-square law to date come from very precise measurements of the Moon's orbit about the Earth. Even though the Moon's orbit has a mean radius of 384,000 km, the models agree with the data at the level of 4 mm! As a result, analysis of the LLR data tests the gravitational inverse-square law to $3\times10^{-11}$ of the gravitational field strength on scales of the Earth-Moon distance \cite{Williams-Turyshev-Murphy-2004}.

Additionally, interplanetary laser ranging could provide the conditions needed to improve the tests of the inverse-square law on interplanetary scales \cite{Turyshev-Williams-2007}. MLR could be used to perform an experiment that could reach the accuracy of $1\times 10^{-14}$ at 2-AU distances, thereby improving the current tests  by several orders of magnitude. 

Although most modern experiments do not show disagreements with Newton's law, there are puzzles that require further investigation. The radiometric tracking data received from the Pioneer 10 and -11 spacecraft at heliocentric distances between 20 and 70 AU has consistently indicated the presence of a small, anomalous, Doppler drift in the spacecraft carrier frequency. The drift can be interpreted as arising from a constant sunward acceleration of $a_{\rm P}  = (8.74 \pm 1.33)\times 10^{-10}~{\rm m/s}^2$ for each particular craft \cite{pioprl,pioprd,moriond,problem_set_05}. This apparent violation of the gravitational inverse-square law has become known as the Pioneer anomaly.

The possibility that the anomalous behavior will continue to defy attempts at conventional explanation has resulted in a growing discussion about the origin of the discovered effect, including suggestions for new physics mechanisms \cite{Milgrom01,FootVolkas01,Bertolami:2003ui,orfeu_jorge05,Bekenstein:2004ne,Jaekel-Reynaud:2005,Bruneton:2007si} and proposals for a dedicated deep space experiment \cite{MMNietoTuryshev04,Turyshev:2004zp,ESLAB2005_Pioneer}.\footnote{For details, please consult the web-page of the Pioneer Explorer Collaboration at the International Space Science Institute (ISSI), Bern, Switzerland, {\tt http://www.issi.unibe.ch/teams/Pioneer/}} A recently initiated investigation of the anomalous signal using the entire record of the Pioneer spacecraft telemetry files, in conjunction with the analysis of much-extended Pioneer Doppler data, may soon reveal the origin of the  anomaly \cite{pio-data-Turyshev:2005,pio-data-Toth:2006,pio-data-Toth:2007}. 

Besides the Pioneer anomaly, there are other intriguing puzzles in the solar system dynamics still awaiting a proper explanation, notably the so-called `fly-by anomaly' \cite{fly-by-Anderson:2006,Laemmerzahl-Preuss-Dittus-2006,fly-by-Anderson:2008}, that occurred during Earth gravity assists performed by several interplanetary spacecraft.

\subsection{\label{sec:mod-grav}Tests of Alternative and Modified-Gravity Theories with Gravitational Experiments in Space}
\label{sec:mod-grav-tests}

Given the immense challenge posed by the unexpected discovery of the accelerated expansion of the universe, it is important to explore every option to explain and probe the underlying physics.  Theoretical efforts in this area offer a rich spectrum of new ideas, some discussed below, that can be tested by experiment.

Motivated by the dark-energy and dark-matter problems, long-distance gravity modification is one of the radical proposals that have recently gained attention \cite{Deffayet-Dvali-Gabadadze-2002}.  Theories that modify gravity at cosmological distances exhibit a strong coupling phenomenon of extra graviton polarizations \cite{Deffayet-etal-2002,Dvali-2006}. This phenomenon plays an important role in this class of theories in allowing them to agree with solar system constraints.  In particular, the ``brane-induced gravity'' model \cite{Dvali-Gabadadze-Porrati-2000} provides a new and interesting way of modifying gravity at large distances to produce an accelerated expansion of the universe, without the need for a non-vanishing cosmological constant \cite{Deffayet:2000uy,Deffayet-Dvali-Gabadadze-2002}. One of the peculiarities of this model is means of recovering the usual gravitational interaction at small (i.e. non-cosmological) distances, motivating precision tests of gravity on solar system scales \cite{Bekenstein:2006fi,Magueijo-Bekenstein-2007}.   

The Eddington parameter $\gamma$, whose value in general relativity is unity, is perhaps the most fundamental PPN parameter \cite{Will_book93,Will-lrr-2006-3}, in that $\frac{1}{2}\bar\gamma$ is a measure, for example, of the fractional strength of the scalar-gravity interaction in scalar-tensor theories of gravity \cite{Damour_Nordtvedt_93b}.  Currently, the most precise value for this parameter, $\bar\gamma = (2.1\pm2.3)\times10^{-5}$, was obtained using radio-metric tracking data received from the Cassini spacecraft \cite{Bertotti-Iess-Tortora-2003} during a solar conjunction experiment.\footnote{A similar experiment is planned for the ESA's {BepiColombo} mission to Mercury \cite{Iess-Asmar:2007}.} This accuracy approaches the region where multiple tensor-scalar gravity models, consistent with the recent cosmological observations \cite{Spergel-etal-2006}, predict a lower bound for the present value of this parameter at the level of $\bar\gamma \sim 10^{-6}-10^{-7}$ \cite{Damour_Nordtvedt_93a,Damour_Nordtvedt_93b,Damour_Polyakov_94a,Damour_Polyakov_94b,Damour_Piazza_Veneziano_02a,Damour_Piazza_Veneziano_02b,Damour-EspFarese-1996-2}. Therefore, improving the measurement of this parameter\footnote{In addition, any experiment pushing the present upper bound on another Eddington  parameter $\beta$, i.e. $\beta - 1 = (0.9 \pm 1.1) \times 10^{-4}$ from \cite{Williams-etal-2004,Williams-etal-2005}, will also be of interest.} would provide crucial information to separate modern scalar-tensor theories of gravity from general relativity, probe possible ways for gravity quantization, and test modern theories of cosmological evolution.

The current accuracy of modern optical astrometry, as represented by the Hipparcos Catalogue, is about 1 mas, which gave a determination of $\gamma$ at the level of $0.997 \pm 0.003$ \cite{Froeschle-Mignard-Arenou-1997}.
Future astrometric missions such as the Space Interferometry Mission (SIM) and especially Gaia will push the accuracy to the level of a few microarcseconds, and the expected accuracy of determinations of $\gamma$ will be $10^{-6}$ to $5\times10^{-7}$ \cite{Turyshev-sim:2008}. 

Interplanetary laser ranging could lead to a significant improvement in the accuracy of the parameter $\gamma$. Thus, precision ranging between the Earth and a lander on Mars during solar conjunctions may offer a suitable opportunity (i.e., MLR).\footnote{In addition to Mars, a Mercury lander \cite{Anderson-etal-1997} equipped with a laser ranging transponder would be very interesting as it would probe a stronger gravity regime while providing measurements that will not be affected by the dynamical noise from the asteroids \cite{Ken_asteroids97,Konopliv_etal_2005}.} If the lander were equipped with a laser transponder capable of reaching a precision of 1~mm, a measurement of $\gamma$ with accuracy of few parts in 10$^7$ would be possible. To reach accuracies beyond this level one must rely on a dedicated space experiment \cite{Turyshev-etal-2007,Turyshev-Williams-2007}.

\begin{wrapfigure}{R}{0.18\textwidth}
  \vspace{-10pt}
  \begin{center}
    \includegraphics[width=0.18\textwidth]{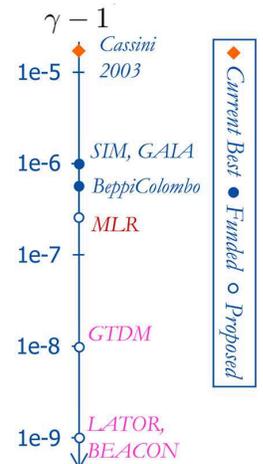}
  \end{center}
  \vspace{-20pt}
  \caption{Anticipated progress in the tests of the PPN parameter $\gamma$ \cite{Turyshev-etal-2007,Turyshev:2008dr}.}
  \vspace{-5pt}
\label{fig:future-gamma-tests}
\end{wrapfigure}

The Gravitational Time Delay Mission (GTDM) \cite{Ashby-Bender-2006,Bender-etal-2006} proposes to use laser ranging between two drag-free spacecraft (with spurious acceleration levels below $1.3 \times  10^{-13}~{\rm m/s}^2/\sqrt{\rm Hz}$ at  $0.4~\mu$Hz) to accurately measure the Shapiro time delay for laser beams passing near the Sun. One spacecraft will be kept at the L1 Lagrange point of the Earth-Sun system; the other one will be  placed on a 3:2 Earth-resonant, LATOR-type, orbit (see paragraph below and \cite{Turyshev-etal-2006,stanford_ijmpd,LATOR-Q2C06} for details). A high-stability frequency standard ($\delta f/f\lesssim 1 \times  10^{-13}~1/\sqrt{\rm Hz}$ at $0.4~\mu$Hz) located on the L1 spacecraft will permit accurate measurement of the time delay. If requirements on the performance of the disturbance compensation system, the timing-transfer process, and the high-accuracy orbit determination are successfully addressed \cite{Bender-etal-2006}, then determination of the time delay of interplanetary signals to a 0.5 ps precision in terms of the instantaneous clock frequency could lead to an accuracy of 2 parts in $10^{8}$ in measuring parameter $\gamma$.

The Laser Astrometric Test of Relativity (LATOR) \cite{Turyshev-etal-2006,Plowman-Hellings:2005,stanford_ijmpd,LATOR-Q2C06} proposes to measure parameter $\gamma$ with an accuracy of 1 part in 10$^9$, which is a factor of 30000 beyond the  best currently available, Cassini's 2003 result \cite{Bertotti-Iess-Tortora-2003}.  The key element of LATOR is a geometric redundancy provided by the long-baseline optical interferometry and interplanetary laser ranging. By using a combination of independent time-series of gravitational deflection of light in immediate proximity to the Sun, along with measurements of the Shapiro time delay on interplanetary scales (to a precision better than 0.01 picoradians and 3 mm, respectively), LATOR will significantly improve our knowledge of relativistic gravity and cosmology. LATOR's primary measurement, the precise observation of the non-Euclidean geometry of a light triangle that surrounds the Sun, pushes to unprecedented accuracy the search for cosmologically relevant scalar-tensor theories of gravity by looking for a remnant scalar field in today's solar system.  LATOR could lead to very robust advances in the tests of fundamental physics. It could discover a violation or extension of general relativity or reveal the presence of an additional long range interaction.

Similar to LATOR, the Beyond Einstein Advanced Coherent Optical Network (BEACON) \cite{Turyshev-etal-2007:BEACON,Turyshev-etal-2008:BEACON} is an experiment designed to reach a sensitivity of 1 part in 10$^9$ in measuring the PPN parameter $\gamma$. The mission will place four small spacecraft in 80,000 km circular orbits around the Earth with all spacecraft in the same plane.  Each spacecraft will be equipped with three sets of identical laser ranging transceivers, which will send laser metrology beams between the spacecraft to form a flexible light-trapezoid formation.  In Euclidean geometry this system is redundant. By measuring only five of the six distances one can compute the sixth.  To enable its primary science objective, BEACON will precisely measure and monitor all six inter-spacecraft distances within the trapezoid using transceivers capable of reaching an accuracy of $\sim0.1$~nm in measuring these distances. The resulting geometric redundancy is the key element that enables BEACON's superior sensitivity in measuring a departure from Euclidean geometry.  In the Earth's vicinity, this departure is primarily due to the curvature of the relativistic space-time. It amounts to $\sim10$~cm for laser beams just grazing the surface of the Earth, then falls off inversely proportional to the impact parameter.  Simultaneous analysis of the resulting time-series of these distance measurements will allow BEACON to measure the curvature of the space-time around the Earth with an accuracy of better than 1 part in $10^9$ \cite{Turyshev-etal-2007:BEACON}.    

\section{Discussion and Outlook}
\label{sec:conclusions}

Today physics stands at the threshold of major discoveries.  Growing observational evidence points to the need for new physics. As a result, efforts to discover new fundamental symmetries, investigations of the limits of established symmetries, tests of the general theory of relativity, searches for gravitational waves, and attempts to understand the nature of dark matter and dark energy are among the main research topics in fundamental physics today \cite{Turyshev-etal-2007,Turyshev:2008jd}. 

The remarkable recent progress in observational cosmology has subjected the general theory of relativity to increased scrutiny by suggesting a non-Einsteinian scenario of the Universe's evolution. From a theoretical standpoint, the challenge is even stronger---if gravity is to be quantized, general relativity must be modified. Furthermore, recent advances in the scalar-tensor extensions of gravity, and brane-world gravitational models, along with efforts to modify gravity on large scales, are motivating new searches for experimental signatures of very small deviations from general relativity on various scales, including on the spacecraft-accessible distances in the solar system. These theoretical advances are motivating searches for very small deviations from Einstein's theory, at the level of three to five orders of magnitude below the level currently tested by experiment.

This progress was matched by the major improvements in measurement technologies. Today, a new generation of high-performance quantum sensors (e.g., ultrastable atomic clocks, accelerometers, gyroscopes, gravimeters, gravity gradiometers) is surpassing previous state-of-the-art instruments, demonstrating the high potential of these techniques based on the engineering and manipulation of atomic systems. Atomic clocks and inertial quantum sensors represent a key technology for accurate frequency measurements and ultraprecise monitoring of accelerations and rotations (see discussion in Ref.~\cite{Turyshev-etal-2007}). 
New quantum devices based on ultra-cold atoms will enable fundamental physics experiments testing quantum physics, physics beyond the Standard Model of fundamental particles and interactions, special relativity, gravitation and general relativity \cite{Tino-etal:2007}. 

The experiments presented here are just few examples of the rich opportunities available to explore the nature of the physical laws. Together with the ground-based laboratories (i.e., such as Large Hadron Collider\footnote{CERN's Large Hadron Collider website {\tt http://lhc.web.cern.ch/lhc/}} at CERN), these space-based experiments could potentially offer an improvement of up to several orders of magnitude over the accuracy of current tests. If implemented, the missions discussed above could significantly advance research in fundamental physics. This progress promises very important new results in gravitational research over the next decade.

\section*{Acknowledgments}
I would like to express my gratitude to my colleagues who have either collaborated with me on this manuscript or given me their wisdom.  I specifically thank Eric G. Adelberger, Peter L. Bender, Philippe Bouyer, Luigi Cacciapuoti, Anatoly M. Cherepaschuk, Curt J. Cutler, Thibault Damour, Gia Dvali, Harvey Gould, John L. Hall, V. Alan Kosteleck\'y, Alex Kusenko, Claus L\"ammerzahl, Kenneth L. Nordtvedt, Craig Hogan, William D. Phillips, Robert D. Reasenberg, Serge Reynaud, Albrecht Ruediger, Christophe Salomon, Irwin I. Shapiro, Guglielmo Tino, Jun Ye, Frank Wilczek  and  Vladimir E. Zharov who provided me with valuable comments, encouragement, support and stimulating discussions while this paper was in preparation.

The work described here was carried out the Jet Propulsion Laboratory, California Institute of Technology, Pasadena, California, under a contract with the National Aeronautics and Space Administration and at the Sternberg Astronomical Institute, Moscow State University, Moscow, Russia.
  
\bibliography{gr-tests}

\end{document}